\documentclass[letterpaper,journal]{IEEEtran} 
\usepackage{amsmath,amsfonts}
\usepackage{amssymb}
\usepackage{array}
\usepackage{textcomp}
\usepackage{stfloats}
\usepackage{url}
\usepackage{verbatim}
\usepackage{comment} 
\usepackage{bm} 
\def\BibTeX{{\rm B\kern-.05em{\sc i\kern-.025em b}\kern-.08em
    T\kern-.1667em\lower.7ex\hbox{E}\kern-.125emX}}
\usepackage{balance}
\usepackage{graphicx} 
\usepackage{subcaption}  
\usepackage{algorithm} 
\usepackage{algpseudocode} 
\usepackage{xcolor}
\begin{document}
\title{Sensing-Assisted Secure Communication in MA-Aided ISAC: CRB Analysis and Robust Design}
\author{Yaxuan Chen, Guangchi Zhang, Miao Cui, Hao Fu, Qingqing Wu, \IEEEmembership{Senior Member, IEEE}, \\  and Rui Zhang, \IEEEmembership{Fellow, IEEE}
\thanks{This work of G. Zhang and M. Cui was supported by Guangdong Basic and Applied Basic Research Foundation under Grant 2026A1515011208. This work of Q. Wu was supported by NSFC 62371289 and NSFC 62331022. 

Yaxuan Chen, Guangchi Zhang, Miao Cui, Hao Fu are with the School of Information Engineering, Guangdong University of Technology, Guangzhou 510006, China (e-mail: chenyaxuancyx@126.com; \{gczhang, cuimiao, ecejasonhaofu\}@gdut.edu.cn).

Qingqing Wu is with the Department of Electronic Engineering, Shanghai Jiao Tong University, Shanghai 200240, China (e-mail: qingqingwu@sjtu.edu.cn).

Rui Zhang is with the Department of Electrical and Computer Engineering, National University of Singapore, Singapore 117583, Singapore (e-mail: elezhang@nus.edu.sg).

Corresponding author: G. Zhang.}
}

\maketitle 

\begin{abstract}
A core challenge in physical-layer security is the difficulty of obtaining the channel state information (CSI) of potential eavesdroppers. The inherent sensing functionality of integrated sensing and communication (ISAC) systems offers a promising solution by enabling the estimation of key parameters, such as the eavesdropper's angles of departure (AoDs). Capitalizing on this capability, we propose a sensing-assisted secure communication scheme for a movable antenna (MA)-aided ISAC system. The scheme comprises two stages: eavesdropper AoD sensing and secure communication. In the first stage, the base station (BS) optimizes the positions of its transmit and receive MAs to enhance sensing accuracy. We derive the closed-form Cramér-Rao bound (CRB) for the estimated AoDs to fundamentally characterize how MA positions influence the estimation uncertainty. In the second stage, the BS ensures secure communication by designing a robust beamforming vector that accounts for the AoD uncertainty region and by further optimizing the transmit MAs' positions to maximize the secrecy rate. To manage the end-to-end design, we formulate a joint optimization problem. This intractable non-convex problem is decomposed into two subproblems. For the first subproblem, we develop an alternating optimization (AO) algorithm to solve the CRB minimization problem. For the second subproblem, we solve the worst-case secrecy rate maximization problem using a method based on backward induction, convex hull construction, and AO. Finally, simulation results are provided to demonstrate the significant advantages of the proposed scheme compared to various benchmarks.
\end{abstract}


\section{Introduction}
\IEEEPARstart{T}{he} evolution towards next-generation wireless networks has engendered a widespread reliance on wireless information transmission, concurrently elevating the importance of robust data security and privacy \cite{RF1-1}. This challenge is compounded by the limitations of traditional upper-layer encryption, which often entails high computational complexity and significant key management overhead \cite{RF1-2}. In this context, physical layer security (PLS) has emerged as a compelling alternative, leveraging the inherent randomness of the physical medium and the disparity between legitimate and wiretap channels to secure communications \cite{RF1-3}. However, a major impediment to the practical deployment of PLS is its frequent reliance on the idealistic assumption of perfect eavesdropper channel state information (CSI), a condition rarely met for passive or unauthorized eavesdroppers.

Fortunately, the innate sensing functionality of integrated sensing and communication (ISAC) systems offers a viable solution to this challenge \cite{RF1-4}. ISAC represents a new paradigm where sensing and communication are not merely coexistent but mutually beneficial, yielding advantages such as enhanced security, higher spectral efficiency, and improved energy efficiency \cite{RF1-5}. This synergy is particularly evident in security applications. For example, the work in \cite{RF1-6} leveraged ISAC sensing to inform secure transmission strategies, including a covert communication scheme that anticipates adversary behavior \cite{RF1-7} and a sensing-assisted PLS scheme that dynamically adapts security measures based on inferred eavesdropper locations \cite{RF1-8}. Inspired by these advancements, this work explores the benefits of integrating sensing into secure communication design.

To suppress signal reception at eavesdroppers while enhancing it at legitimate receivers, multiple-antenna techniques that exploit spatial degrees of freedom (DoFs) have been widely adopted for PLS in ISAC systems \cite{RF1-9}. A key application is robust beamforming, which has shown significant promise in addressing eavesdropper channel uncertainty \cite{RF1-10}-\cite{RF1-12}. The authors of \cite{RF1-10} formulated beampattern optimization problems under both bounded and Gaussian CSI error models. Similarly, \cite{RF1-11} investigated secure beamforming under perfect and imperfect CSI conditions, while \cite{RF1-12} proposed a robust design based on statistical CSI for both the legitimate and eavesdropping links.

However, the performance of conventional multi-antenna beamforming is fundamentally limited by its reliance on fixed-position antennas (FPAs), which can lead to suboptimal secrecy performance, particularly when the channels of the legitimate receiver and the eavesdropper are highly correlated \cite{RF1-A}, \cite{RF1-B}. To overcome this limitation, movable antenna (MA) technology has emerged as a new paradigm \cite{RF1-13}. By flexibly adjusting antenna positions within a given region, MAs introduce additional spatial DoFs that can be exploited to improve channel conditions. Specifically, MA position optimization can actively mitigate the channel correlation between the legitimate receiver and the eavesdropper, even when they are in similar directions, thereby unlocking the full potential of secure beamforming. Consequently, MA-aided secure communication has attracted growing research interest \cite{RF1-14}-\cite{RF1-16}. While these works have demonstrated the potential of MAs, they often depend on the ideal assumption of perfect eavesdropper CSI. More recent studies have begun to address this by considering Rician fading channels with known line-of-sight (LoS) components \cite{RF1-17} or by modeling location uncertainty \cite{RF1-18}.

Despite these advances, critical gaps remain. The aforementioned works have not fully explored the potential of actively sensing the eavesdropper's direction to reduce channel acquisition overhead and enhance security. Furthermore, most research on MA-aided ISAC systems treats sensing and communication as separate objectives requiring a trade-off \cite{RF1-19}, \cite{RF1-20}, rather than as symbiotic functions for security enhancement. While the benefits of MAs in ISAC for improving spectral efficiency and beamforming flexibility are well-documented \cite{RF1-21}, \cite{RF1-22}, research on PLS in MA-aided ISAC systems remains nascent. The few existing works, such as \cite{RF1-23} and \cite{RF1-24}, operate under the strong assumption of perfect eavesdropper CSI obtained through error-free sensing. In practice, however, sensing errors are non-negligible. Moreover, in MA-aided ISAC systems, the sensing accuracy depends on the MAs’ positions. These positions directly determine the estimation uncertainty region of the eavesdropper’s channel, and consequently affect the achievable secrecy performance. How to explicitly characterize this estimation uncertainty region and actively account for it to guarantee secure communication remains a largely unexplored challenge. 

To address this gap, we propose a sensing-assisted secure communication scheme that incorporates the AoD uncertainty region characterized by the Cramér-Rao Bound (CRB) into the robust beamforming design, thereby improving secrecy rate in MA-aided ISAC systems. The main contributions of this work are summarized as follows:
\begin{itemize}
    \item We propose a two-stage sensing-assisted secure communication scheme for an MA-aided ISAC system. In the first stage, the base station (BS) senses the eavesdropper's angles of departure (AoDs) by optimizing the positions of its transmit and receive MAs ($\tilde{\mathbf{t}}_s$ and $\tilde{\mathbf{r}}$) to enhance sensing accuracy. In the second stage, the resulting AoD uncertainty region is explicitly considered in the secure communication stage, where the robust beamforming vector ($\boldsymbol{\omega}$) and the transmit MAs' positions ($\tilde{\mathbf{t}}_c$) are jointly optimized to maximize the security performance.
    \item In the eavesdropper sensing stage, we derive a closed-form expression for the CRB of the AoD estimates as a function of the transmit and receive MAs' positions, explicitly characterizing the relationship between the MA geometry and the achievable estimation accuracy. To reduce the AoD uncertainty region, we formulate and solve a CRB minimization problem by jointly optimizing $\tilde{\mathbf{t}}_s$ and $\tilde{\mathbf{r}}$ using an alternating optimization (AO) algorithm.
    \item In the secure communication stage, we formulate a worst-case secrecy rate maximization problem that strictly accounts for the AoD uncertainty region characterized by the CRB. Unlike conventional robust beamforming designs with FPAs, this joint optimization of $\tilde{\mathbf{t}}_c$ and $\boldsymbol{\omega}$ exploits the additional spatial DoFs provided by MAs to improve secrecy rate under estimation uncertainty region. We solve this non-convex problem using backward induction, where for given transmit MAs' positions, the robust beamforming is obtained via convex hull construction, and the MAs' positions are then optimized by using an AO algorithm.
    \item We provide extensive simulation results to validate the advantages of the proposed scheme. The results confirm the significant CRB reduction achieved by MAs over FPAs and demonstrate the performance benefits of robust design within the AoD uncertainty region. A comparative analysis highlights the critical role of sensing in boosting the secrecy rate and the superior robustness of our scheme against benchmarks that neglect estimation uncertainty.
\end{itemize}

The remainder of this paper is organized as follows. Section II presents the system model. Section III develops the CRB analysis and formulates the optimization problem. Section IV details the MA position optimization for the sensing stage, followed by the joint design for the secure communication stage in Section V. Finally, Section VI provides simulation results, and Section VII concludes the paper.

\textit{Notations}: Vectors and matrices are denoted by bold lowercase and uppercase letters, respectively. The conjugate, transpose, and conjugate transpose are represented by $(\cdot)^*$, $(\cdot)^T$, and $(\cdot)^H$. diag$\{\cdot\}$ denotes the diagonalization operation. $\mathbb{R}^{A \times B}$ and $\mathbb{C}^{A \times B}$ represent the sets of $A \times B$ real and complex matrices. $\mathbf{I}_A$ is the $A \times A$ identity matrix, and $\mathbf{1}_A$ is an $A \times 1$ vector of ones. The $p$-th entry of a vector $\mathbf a$ is $\mathbf a[p]$, and the $(p,q)$-th entry of a matrix $\mathbf A$ is $\mathbf A[p, q]$. The $2$-norm of a vector $\mathbf a$ is $||\mathbf a||_2$. vec($\mathbf{A}$) is the vectorization of matrix $\mathbf{A}$. $\otimes$ is the Kronecker product. $\Re\{a\}$ and $\Im\{a\}$ denote the real and imaginary parts of a complex number $a$.

\begin{figure*}[!t]
    \centering
    \begin{minipage}{0.495\textwidth}
        \centering
        \includegraphics[width=\linewidth]{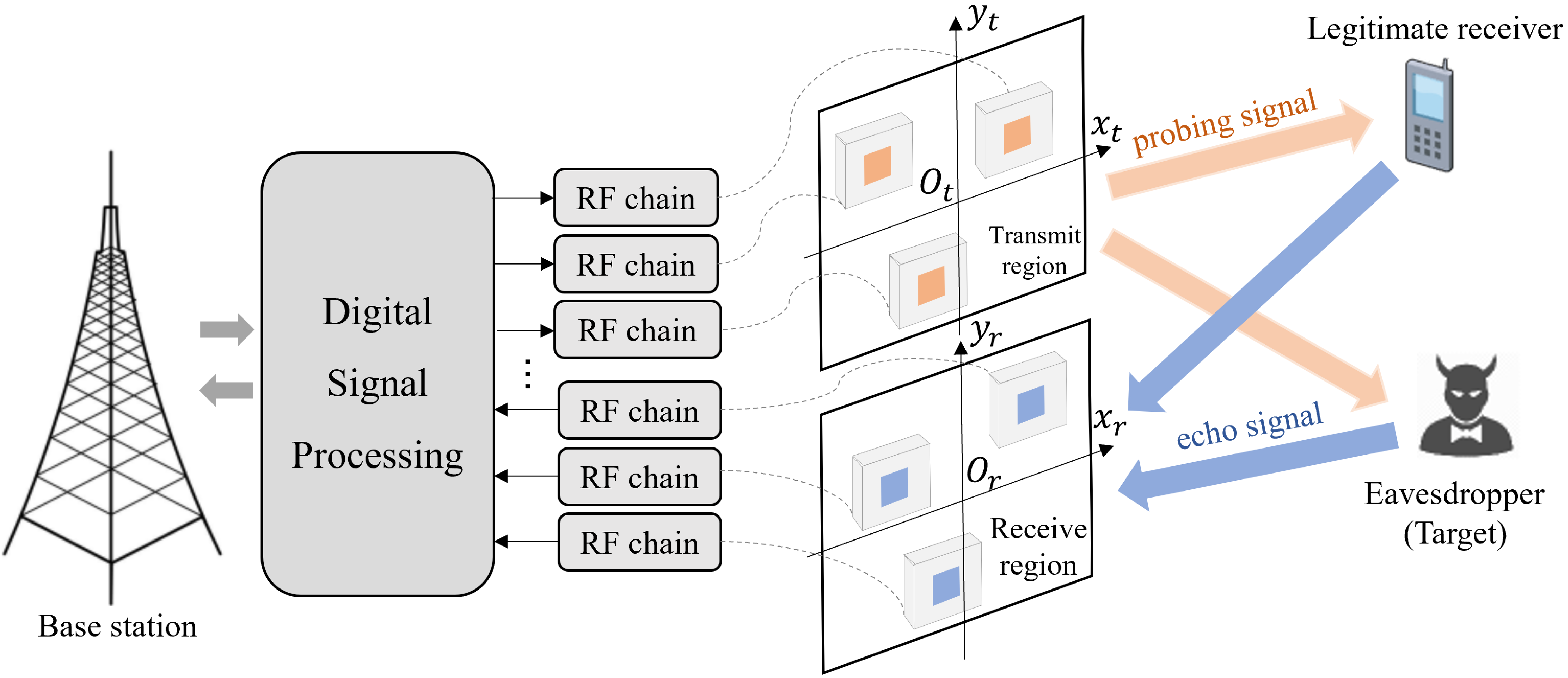}
        \par\smallskip 
        (a) Eavesdropper sensing stage. 
        \label{fig:sub1}
    \end{minipage}
    \hfill
    \begin{minipage}{0.495\textwidth}
        \centering
        \includegraphics[width=\linewidth]{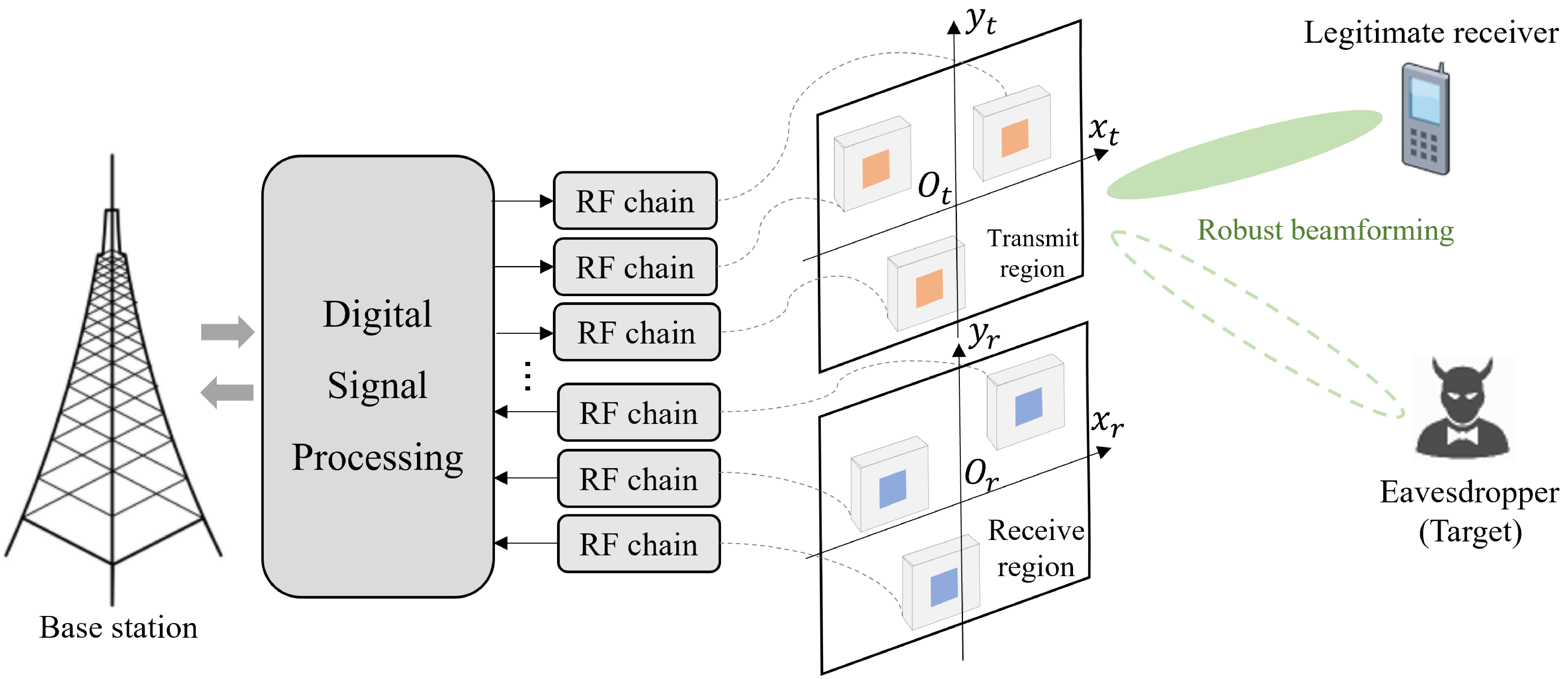}
         (b) Secure communication stage. 
        \label{fig:sub2}
    \end{minipage}
    \caption{An MA-aided ISAC system.}
    \label{fig:main}
\end{figure*}

\begin{figure}[!t]
    \centering
    \begin{minipage}{0.24\textwidth}
        \centering
        \includegraphics[width=\textwidth]{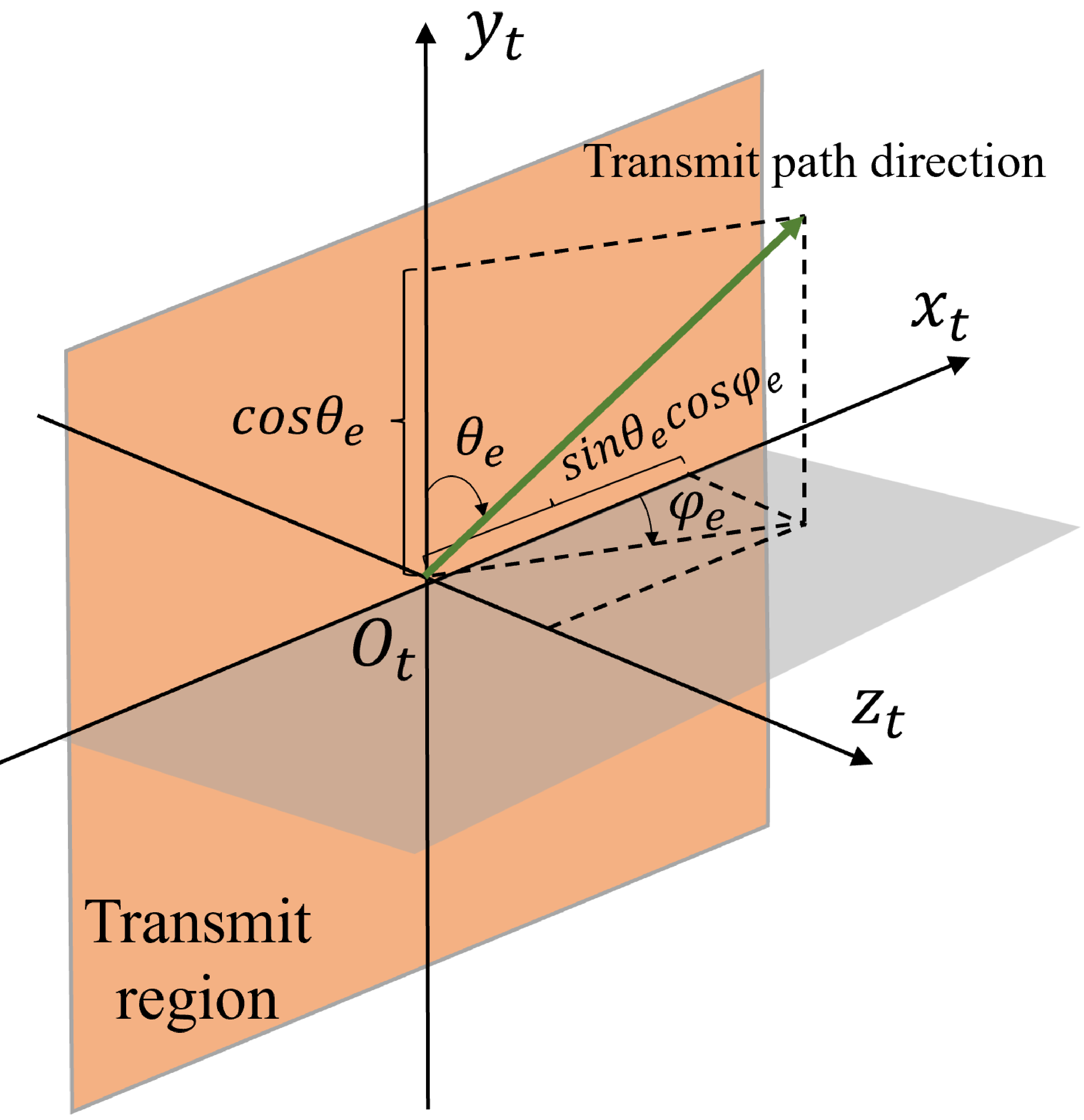}
        \par\smallskip
        (a) Transmit MA region.
        \label{fig:sub3}
    \end{minipage}
    \hfill
    \begin{minipage}{0.24\textwidth}
        \centering
        \includegraphics[width=\textwidth]{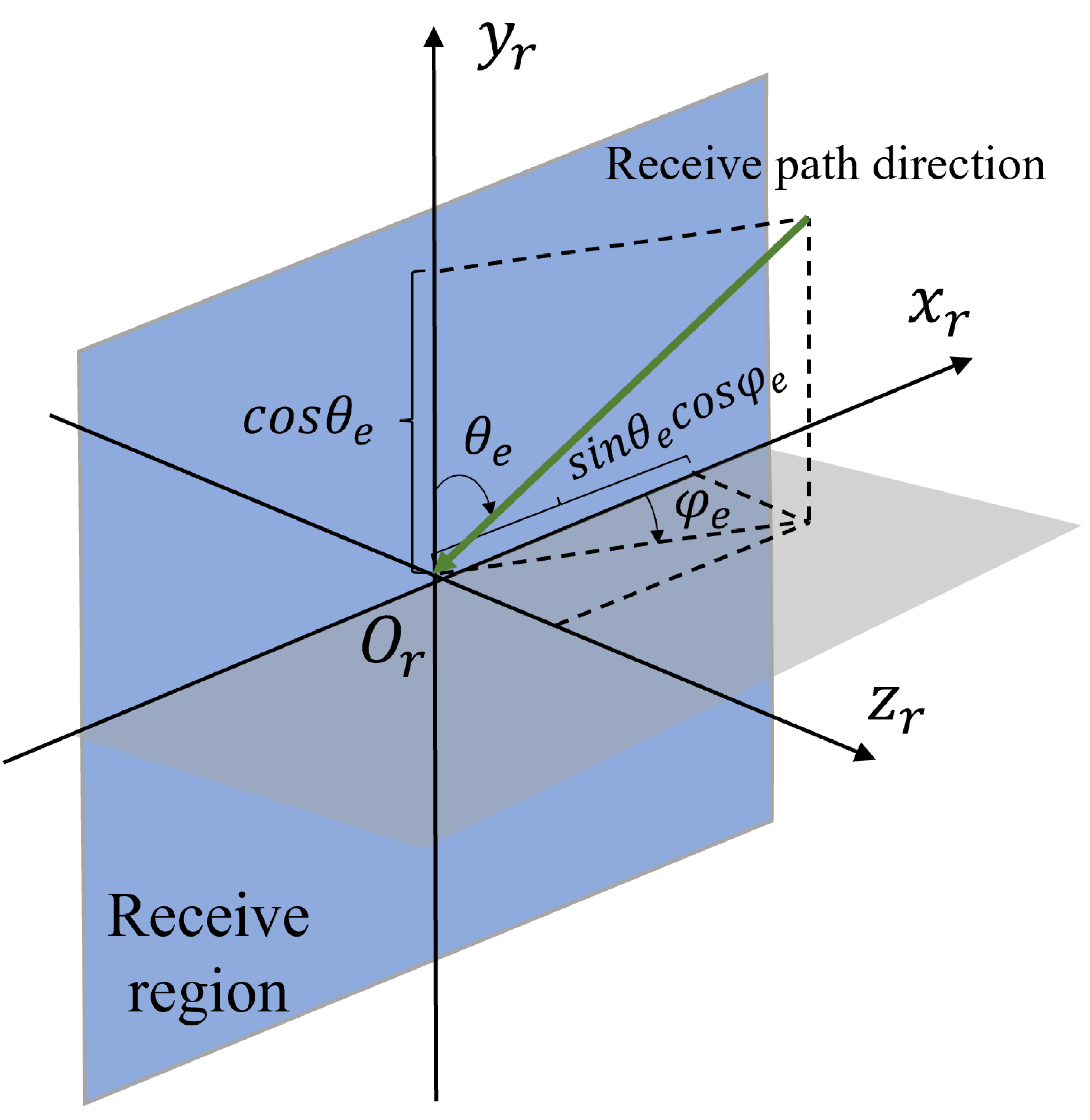}
        \par\smallskip
        (b) Receive MA region.
        \label{fig:sub4}
    \end{minipage}
    \caption{Illustrations of the coordinates and spatial angles for the transmit and receive MA regions.}
    \label{fig:picture}
\end{figure}

\section{System Model}
As illustrated in Fig. 1, we consider an MA-aided ISAC system operating in the millimeter-wave (mmWave) band, where a BS performs two main functions: sensing the direction of a potential eavesdropper and securely transmitting information to a legitimate receiver\footnote{For analytical tractability, we focus on a single-eavesdropper scenario. This setup allows us to explicitly reveal the relationship among MA position optimization, sensing accuracy characterized by the CRB, and the resulting secrecy rate.}. The security measures are adapted based on the sensing results. The BS is equipped with an $N$-element transmit MA array and an $M$-element receive MA array\footnote{We assume self-interference (SI) is effectively suppressed using established multi-stage cancellation techniques (e.g., spatial isolation, analog and digital cancellation) \cite{RF1-8}. Any residual SI is modeled as an additive Gaussian random variable and absorbed into the background noise. As a result, the adopted signal model and subsequent analytical results remain valid.}. Both arrays are connected to radio frequency (RF) chains via flexible cables, allowing the antennas to move freely within their respective two-dimensional (2D) regions.

To precisely define the antenna positions, we establish two local 2D Cartesian coordinate systems. The position of the $n$-th transmit MA is denoted by $\mathbf{t}_n=[x_n^t,y_n^t]^T \in \mathcal{C}_t$ for $1\leq n\leq N$, and the position of the $m$-th receive MA is $\mathbf{r}_m=[x_m^r,y_m^r]^T \in \mathcal{C}_r$ for $1\leq m\leq M$. Here, $\mathcal{C}_t$ and $\mathcal{C}_r$ represent the designated movement regions for the transmit and receive MAs, respectively, which we assume to be square regions of size $A \times A$ without loss of generality. We define the collective position vectors for all transmit and receive MAs as $\tilde{\mathbf{t}}\ \triangleq [\mathbf{t}_1^T,\ldots,\mathbf{t}_N^T\ ]^T \in \mathbb{R}^{2N \times 1}$ and $\tilde{\mathbf{r}}\ \triangleq [\mathbf{r}_1^T,\ldots,\mathbf{r}_M^T\ ]^T \in \mathbb{R}^{2M \times 1}$. Both the legitimate receiver and the eavesdropper are equipped with a single FPA.

We assume the BS has perfect CSI of the legitimate receiver but, in a more practical scenario, lacks any prior CSI of the eavesdropper. To overcome this, we propose a sensing-assisted secure scheme that divides each transmission process into two stages: an eavesdropper sensing stage followed by a secure communication stage. In the first stage, the BS transmits probing signals and receives the corresponding echoes to estimate the eavesdropper's angles of departure (AoDs)\footnote{In a monostatic sensing setup, the eavesdropper's angles of arrival (AoAs) at the BS (Fig. 2(b)) are identical to its AoDs from the BS \cite{RF2-3}. Therefore, we focus on estimating the AoDs.}. The transmit and receive MAs are strategically positioned to optimize sensing performance. In the second stage, using the estimated AoD range, the BS transmits confidential data to the legitimate receiver. It employs a robust beamforming vector and re-optimizes the transmit MA positions to minimize information leakage to the eavesdropper. The details of these two stages are described below.

\subsection{Eavesdropper Sensing Stage}
In this stage, the BS transmits probing signals and processes the echoes reflected by the legitimate receiver and the eavesdropper. By canceling the known echo component from the legitimate receiver, the BS can isolate the eavesdropper's echo and estimate its AoDs. The positions of the transmit and receive MAs are optimized to enhance the accuracy of this estimation.

Since the line-of-sight (LoS) path typically exhibits a stronger channel gain than non-line-of-sight (NLoS) components in mmWave systems, we model the BS-eavesdropper and eavesdropper-BS links as LoS channels, consistent with \cite{RF2-1} and \cite{RF2-C}. As shown in Fig. 2(a), the eavesdropper's elevation and azimuth AoDs are denoted by $\theta_e \in [0,\pi]$ and $\varphi_e \in [0,\pi]$, respectively. The corresponding spatial AoDs with respect to (w.r.t.) the $x_t$ and $y_t$ axes are defined as
\begin{equation}\label{1}
\alpha_e\triangleq\sin{\theta_e}\cos{\varphi_e},\;  \beta_e\triangleq\cos{\theta_e}.
\end{equation}
Let $\bm{\rho}_e\triangleq[\alpha_e,\beta_e]^T$ be the 2D normalized wavevector for the BS-eavesdropper channel. The propagation distance difference for the $n$-th transmit MA at position $\mathbf{t}_n$, relative to the origin $O_t$ of the transmit region, is $\psi_e^t(\mathbf{t}_n,\bm{\rho}_e)=x_n^t\alpha_e+y_n^t\beta_e=\bm{\rho}_e^T\mathbf{t}_n$. With $\lambda$ as the carrier wavelength, the resulting phase difference is $\frac{2\pi}{\lambda}\psi_e^t(\mathbf{t}_n,\bm{\rho}_e)$. The field-response vector (FRV) of the transmit MA array for the eavesdropper is then given by \cite{RF1-13}
\begin{equation}\label{2}
\mathbf{g}_e (\tilde{\mathbf{t}},\bm{\rho}_e)=[e^{j\frac{2\pi}{\lambda}\psi_e^t (\mathbf{t}_1,\bm{\rho}_e)},\ldots,e^{j\frac{2\pi}{\lambda}\psi_e^t(\mathbf{t}_N,\bm{\rho}_e)}]^T \in \mathbb{C}^{N \times 1}.
\end{equation}
Similarly, the FRV of the receive MA array for the eavesdropper is
\begin{equation}\label{3}
\mathbf{f}_e (\tilde{\mathbf{r}},\bm{\rho}_e)=[e^{j\frac{2\pi}{\lambda}\psi_e^r (\mathbf{r}_1,\bm{\rho}_e)},\ldots,e^{j\frac{2\pi}{\lambda}\psi_e^r(\mathbf{r}_M,\bm{\rho}_e)}]^T \in \mathbb{C}^{M \times 1},
\end{equation}
where $\psi_e^r(\mathbf{r}_m,\bm{\rho}_e)=x_m^r\alpha_e+y_m^r\beta_e=\bm{\rho}_e^T\mathbf{r}_m$ is the distance difference for the $m$-th receive MA at position $\mathbf{r}_m$ relative to the origin $O_r$. Consequently, the round-trip echo channel for the BS-eavesdropper-BS link is modeled as \cite{RF2-C}
\begin{equation}\label{4}
\mathbf{H}_s (\tilde{\mathbf{t}},\tilde{\mathbf{r}},\bm{\rho}_e)=\zeta_s\mathbf{f}_e(\tilde{\mathbf{r}},\bm{\rho}_e)\mathbf{g}_e(\tilde{\mathbf{t}},\bm{\rho}_e)^H,
\end{equation}
where $\zeta_s = \sqrt{\frac{\lambda^2\epsilon}{64\pi^3d_{be}^4}} e^{j\frac{4\pi d_{be}}{\lambda}}$ is the complex path gain, with $\epsilon$ being the radar cross section (RCS) and $d_{be}$ the distance between the BS and the eavesdropper.

We model the potential eavesdropper as a passive object with a non-zero RCS that reflects the incident sensing signals, enabling its spatial parameters to be estimated from the received echoes. For given MA positions $\tilde{\mathbf{t}}$ and $\tilde{\mathbf{r}}$, the BS estimates the eavesdropper's spatial AoDs, $\alpha_e$ and $\beta_e$, using $T$ snapshots of reflected echoes. We assume that the echo corresponding to the eavesdropper can be separated from other signal components by using standard radar signal processing techniques, such as clutter map subtraction, Doppler filtering, and spatial nulling \cite{RF-add}. At the $t$-th snapshot ($1\leq t\leq T$), the received echo signal is
\begin{equation}\label{5}
\mathbf{y}_s(t)= \mathbf{H}_s (\tilde{\mathbf{t}},\tilde{\mathbf{r}},\bm{\rho}_e) \mathbf{x}_s (t)+\mathbf{z}_s (t),
\end{equation}
where $\mathbf{x}_s (t)\in\mathbb{C}^{N \times 1}$ is the probing signal with average power $\mathbb{E}\{||\mathbf{x}_s (t)||^2_2\}=P_s$, and $\mathbf{z}_s(t)$ is an additive white Gaussian noise (AWGN) vector with zero mean and covariance $\sigma_s^2\mathbf{I}_M$. Stacking the received signals from all snapshots, we obtain the matrix form
\begin{equation}\label{6}
\mathbf{Y}_s\triangleq[\mathbf{y}_s(1),\ldots,\mathbf{y}_s(T)] = \mathbf{H}_s (\tilde{\mathbf{t}},\tilde{\mathbf{r}},\bm{\rho}_e) \mathbf{X}_s+\mathbf{Z}_s,
\end{equation}
where $\mathbf{X}_s=[\mathbf{x}_s (1),\ldots,\mathbf{x}_s (T)] \in \mathbb{C}^{N \times T}$ and $\mathbf{Z}_s=[\mathbf{z}_s (1),\ldots,\mathbf{z}_s (T)] \in \mathbb{C}^{M \times T}$. Following \cite{RF2-4}, we assume $\mathbf{X}_s$ is designed to generate an omnidirectional beampattern to scan for the eavesdropper in all directions. The covariance matrix of the probing signals is thus $\mathbf{R}_{\mathbf{X}_s}=\frac{1}{T}\mathbf{X}_s\mathbf{X}_s^H=\frac{P_s}{N}\mathbf{I}_N$. This requires $\mathbf{X}_s$ to be an orthogonal matrix, necessitating $T \geq N$. For instance, $\mathbf{X}_s$ can be formed using the first $N$ columns of a $T \times T$ discrete Fourier transform (DFT) matrix, with entries
\begin{equation}\label{7}
\mathbf{X}_s[n,t] = \sqrt{\frac{P_s}{N}}e^{j\frac{2\pi(n-1)(t-1)}{T}}.
\end{equation}
Based on these received signals, we estimate the spatial AoDs $\alpha_e$ and $\beta_e$ using the maximum likelihood estimation (MLE) method, and we derive the CRB for performance analysis, as detailed in Section III-A.

\subsection{Secure Communication Stage}
Using the estimated angular range for $\bm{\rho}_e$ obtained from the sensing stage, the BS transmits confidential information to the legitimate receiver. To enhance security, the BS employs a robust beamforming vector and jointly optimizes the transmit MAs' positions.

Let the elevation and azimuth AoDs of the legitimate receiver be $\theta_c\in[0,\pi]$ and $\varphi_c\in[0,\pi]$, respectively. The corresponding spatial AoDs w.r.t. the $x_t$ and $y_t$ axes are $\alpha_c\triangleq\sin\theta_c\cos\varphi_c$ and $\beta_c\triangleq\cos\theta_c$. We define the normalized wavevector as $\bm{\rho}_c\triangleq[\alpha_c,\beta_c]^T$. The propagation distance difference for the $n$-th transmit MA is $\psi_c^t(\mathbf{t}_n,\bm{\rho}_c)=x_n^t\alpha_c+y_n^t\beta_c=\bm{\rho}_c^T\mathbf{t}_n$. Thus, the FRV of the transmit MA array for the legitimate receiver is
\begin{equation}\label{8}
\mathbf{g}_c (\tilde{\mathbf{t}},\bm{\rho}_c)=[e^{j\frac{2\pi}{\lambda}\psi_c^t (\mathbf{t}_1,\bm{\rho}_c)},\ldots,e^{j\frac{2\pi}{\lambda}\psi_c^t(\mathbf{t}_N,\bm{\rho}_c)}]^T \in \mathbb{C}^{N \times 1}.
\end{equation}
The channel vector from the BS to the legitimate receiver, $\mathbf{h}_c (\tilde{\mathbf{t}},\bm{\rho}_c)\in\mathbb{C}^{N \times 1}$, is then expressed as \cite{RF2-C}
\begin{equation}\label{9}
\mathbf{h}_c (\tilde{\mathbf{t}},\bm{\rho}_c)=\zeta_c\mathbf{g}_c(\tilde{\mathbf{t}},\bm{\rho}_c),
\end{equation}
where $\zeta_c=\frac{\lambda}{4\pi d_{bc}}e^{j\frac{2\pi d_{bc}}{\lambda}}$ is the complex path gain, with $d_{bc}$ being the BS-receiver distance. The signal received at the legitimate receiver is
\begin{equation}\label{10}
y_c=\mathbf{h}_c^H (\tilde{\mathbf{t}},\bm{\rho}_c)\boldsymbol{\omega}x_c+z_c,
\end{equation}
where $\boldsymbol{\omega} \in \mathbb{C}^{N \times 1}$ is the robust transmit beamforming vector, $x_c$ is the transmitted signal with normalized power, and $z_c \sim \mathcal{CN}(0,\sigma^2_c)$ is the AWGN. Similarly, the signal received at the eavesdropper is
\begin{equation}\label{11}
y_e=\mathbf{h}_e^H (\tilde{\mathbf{t}},\bm{\rho}_e)\boldsymbol{\omega}x_c+z_e,
\end{equation}
where $\mathbf{h}_e (\tilde{\mathbf{t}},\bm{\rho}_e)=\zeta_e\mathbf{g}_e(\tilde{\mathbf{t}},\bm{\rho}_e)$, $\zeta_e=\frac{\lambda}{4\pi d_{be}}e^{j\frac{2\pi d_{be}}{\lambda}}$ is the complex path gain from the BS to the eavesdropper, and $z_e \sim \mathcal{CN}(0,\sigma^2_e)$ is the AWGN.

The achievable rates for the legitimate receiver and the eavesdropper are respectively given by
\begin{align}
R_c(\tilde{\mathbf{t}},\boldsymbol{\omega}) & = \log_2 \left( 1 + \frac{|\mathbf{h}_c^H (\tilde{\mathbf{t}},\bm{\rho}_c)\boldsymbol{\omega}|^2}{\sigma_c^2} \right),  \label{12} \\
R_e(\tilde{\mathbf{t}},\bm{\rho}_e,\boldsymbol{\omega}) & = \log_2 \left( 1 + \frac{|\mathbf{h}_e^H (\tilde{\mathbf{t}},\bm{\rho}_e)\boldsymbol{\omega}|^2}{\sigma_e^2} \right). \label{13}
\end{align}
Defining $[x]^+ \triangleq \max\{0, x\}$, the secrecy rate is expressed as
\begin{equation}\label{14}
R_{\text{sec}}(\tilde{\mathbf{t}},\bm{\rho}_e,\boldsymbol{\omega}) = [R_c(\tilde{\mathbf{t}},\boldsymbol{\omega})-R_e(\tilde{\mathbf{t}},\bm{\rho}_e,\boldsymbol{\omega})]^+.
\end{equation}

\section{Sensing CRB Analysis and Problem Formulation}
In this section, we first establish a theoretical benchmark for the sensing performance of the BS by deriving the CRBs for the estimation of the eavesdropper’s AoDs. Building on this analysis, we then formulate a robust design problem that aims to maximize the worst-case secrecy rate under the derived estimation uncertainty by jointly optimizing the MA positions and the beamforming vector.

\subsection{CRB Analysis for Eavesdropper’s AoD Estimation}
To assess the fundamental limits of AoD estimation accuracy, we derive the CRB, which provides a lower bound on the variance of any unbiased estimator \cite{RF2-5}. Based on the echo signals received at the BS, as modeled in \eqref{6}, we can estimate the eavesdropper's AoDs using the MLE method and subsequently derive their corresponding CRBs.

\textit{\textbf{Lemma 1:} Using the MLE method, the spatial AoDs of the eavesdropper can be estimated as}
\begin{equation}\label{15}
\hat{\bm{\rho}}_e = \arg\max_{\bm{\bar{\rho}}_e\in\Xi_1}  |(\mathbf{f}_e (\tilde{\mathbf{r}},\bm{\bar{\rho}}_e)\otimes\mathbf{g}_e (\tilde{\mathbf{t}},\bm{\bar{\rho}}_e))^H \text{vec}(\mathbf{X}_s(\mathbf{Y}_s)^H )|^2,
\end{equation}
\textit{where the solution can be found by an exhaustive search for $\bm{\bar{\rho}}_e=[\bar{\alpha}_e,\bar{\beta}_e]^T$ over the domain $\Xi_1=[-1,1]\times[-1,1]$}.

\textit{\textbf{Proof:}} See Appendix A. \hfill $\blacksquare$

With the estimator $\hat{\bm{\rho}}_e=[\hat{\alpha}_e,\hat{\beta}_e]^T$ from Lemma 1, the estimation performance can be characterized by the mean squared errors (MSEs), $\text{MSE}_{\alpha_e}=\mathbb{E}\{|\alpha_e-\hat{\alpha}_e|^2\}$ and $\text{MSE}_{\beta_e}=\mathbb{E}\{|\beta_e-\hat{\beta}_e|^2\}$. Since deriving closed-form expressions for the MSEs is difficult, we instead derive the CRBs, which serve as an asymptotically tight lower bound on the MSEs at moderate-to-high sensing signal-to-noise ratios (SNRs).

\textit{\textbf{Lemma 2:} The CRBs for the two estimated spatial AoDs are given by}
\begin{subequations} \label{16}
\begin{align}
\text{CRB}_{\alpha_e}(\tilde{\mathbf{t}},\tilde{\mathbf{r}})=\frac{G}{v(\mathbf{x}_t)+v(\mathbf{x}_r)-\frac{(c(\mathbf{x}_t,\mathbf{y}_t)+c(\mathbf{x}_r,\mathbf{y}_r))^2}{v(\mathbf{y}_t)+v(\mathbf{y}_r)}},\\
\text{CRB}_{\beta_e}(\tilde{\mathbf{t}},\tilde{\mathbf{r}})=\frac{G}{v(\mathbf{y}_t)+v(\mathbf{y}_r)-\frac{(c(\mathbf{x}_t,\mathbf{y}_t)+c(\mathbf{x}_r,\mathbf{y}_r))^2}{v(\mathbf{x}_t)+v(\mathbf{x}_r)}},
\end{align}
\end{subequations}
\textit{where $G=\frac{\lambda^2\sigma_s^2}{8MP_sT\pi^2|\zeta_s|^2}$. The vectors $\mathbf{x}_t=[x_1^t,\ldots,x_N^t]^T\in \mathbb{R}^{N\times1}$ and $\mathbf{y}_t=[y_1^t,\ldots,y_N^t]^T\in \mathbb{R}^{N\times1}$ collect the horizontal and vertical coordinates of the transmit MAs, while $\mathbf{x}_r=[x_1^r,\ldots,x_M^r]^T\in \mathbb{R}^{M\times1}$ and $\mathbf{y}_r=[y_1^r,\ldots,y_M^r]^T\in \mathbb{R}^{M\times1}$ do the same for the receive MAs. The variance function is defined as $v(\mathbf{x}_t)\triangleq\frac{1}{N}\sum_{n=1}^{N}(x^t_n-\mu(\mathbf{x}_t))^2$, with $\mu(\mathbf{x}_t)$ being the mean of vector $\mathbf{x}_t$. Similarly, the covariance function is defined as $c(\mathbf{x}_t,\mathbf{y}_t)\triangleq\frac{1}{N}\sum_{n=1}^{N}(x^t_n-\mu(\mathbf{x}_t))(y^t_n-\mu(\mathbf{y}_t))$.}

\textit{\textbf{Proof:}} See Appendix B. \hfill $\blacksquare$

Lemma 2 reveals that $\text{CRB}_{\alpha_e}(\tilde{\mathbf{t}},\tilde{\mathbf{r}})$ and $\text{CRB}_{\beta_e}(\tilde{\mathbf{t}},\tilde{\mathbf{r}})$ depend directly on the positions of the transmit and receive MAs. The CRBs decrease as the variances of the antenna coordinates ($v(\mathbf{x}_t)$, $v(\mathbf{y}_t)$, etc.) increase and as the magnitudes of the covariances ($c(\mathbf{x}_t,\mathbf{y}_t)$, etc.) decrease. This implies an inherent trade-off: antennas should be placed far apart to maximize variance, yet be arranged symmetrically to minimize covariance. Therefore, optimizing the MA positions to resolve this trade-off is critical for minimizing the CRBs and achieving high-precision sensing.

\textit{\textbf{Remark 1:}} Although the CRB expression in \eqref{16} shares structural similarity with receive-only MA sensing models (e.g., \cite{RF2-13}), it is derived under a different sensing architecture. In \cite{RF2-13}, the CRB depends solely on receive MAs’ positions. In contrast, in the proposed full-duplex ISAC system, both transmit and receive MAs' positions influence the sensing channel and the Fisher information matrix (FIM), making the CRB jointly dependent on their configuration. This joint dependency introduces additional spatial DoFs and motivates the joint MA position optimization developed in this work.

To facilitate a mathematically tractable robust design, we model the probability density functions (PDFs) of the AoD estimation errors as zero-mean Gaussian distributions with variances equal to their respective CRBs \cite{RF1-8}, i.e., $\varepsilon_{\alpha_e} \sim \mathcal{N}(0, \text{CRB}_{\alpha_e}(\tilde{\mathbf{t}},\tilde{\mathbf{r}}))$ and $\varepsilon_{\beta_e} \sim \mathcal{N}(0, \text{CRB}_{\beta_e}(\tilde{\mathbf{t}},\tilde{\mathbf{r}}))$. Under this model\footnote{Under lower sensing SNR conditions where non-asymptotic estimation errors may occur, the AoD uncertainty region can be conservatively enlarged by adopting a larger scaling factor, thereby maintaining a robust design against increased estimation errors.}, the true AoDs of the eavesdropper, ${\bm{\rho}}_e$, will lie within the uncertainty region $\Xi_2=[\hat{\alpha}_e-3\sqrt{\text{CRB}_{\alpha_e}(\tilde{\mathbf{t}},\tilde{\mathbf{r}})},\hat{\alpha}_e+3\sqrt{\text{CRB}_{\alpha_e}(\tilde{\mathbf{t}},\tilde{\mathbf{r}})}] \times [\hat{\beta}_e-3\sqrt{\text{CRB}_{\beta_e}(\tilde{\mathbf{t}},\tilde{\mathbf{r}})},\hat{\beta}_e+3\sqrt{\text{CRB}_{\beta_e}(\tilde{\mathbf{t}},\tilde{\mathbf{r}})}]$ with a probability of approximately 99.73\% \cite{RF2-6}. Since the CRBs are determined by the MA positions, this AoD estimation range is also a function of the antenna geometry. This result reveals that MA position optimization directly controls the size of the AoD uncertainty region, which in turn determines the worst-case eavesdropper channel conditions in the secure communication stage. Therefore, the MAs’ positions offer a new and powerful design dimension to enhance the secrecy rate by reducing estimation uncertainty.

\subsection{Problem Formulation}
Leveraging the CRB analysis, our goal is to maximize the worst-case secrecy rate by jointly designing the sensing and communication stages. In the sensing stage, the transmit and receive MA positions, ${\tilde{\mathbf{t}}}_s$ and $\tilde{\mathbf{r}}$, are optimized to minimize the CRBs, thereby narrowing the AoD estimation range $\Xi_2$. A smaller uncertainty region allows for more precisely targeted robust beamforming \cite{RF2-7}. In the communication stage, given the estimated AoDs $\hat{\bm{\rho}}_e$ and the corresponding range $\Xi_2$, the transmit MA positions ${\tilde{\mathbf{t}}}_c$ and the robust beamforming vector $\boldsymbol{\omega}$ are optimized to maximize the secrecy rate over all possible true AoDs within that range. The overall joint optimization problem can be formulated as
\begin{subequations}  \label{17}
\begin{align}
\max_{{\tilde{\mathbf{t}}}_s,\tilde{\mathbf{r}},{\tilde{\mathbf{t}}}_c,\boldsymbol{\omega}} & \min_{\hat{\bm{\rho}}_e\in\Xi_1,\bm{\rho}_e\in\Xi_2}   R_{\text{sec}}({\tilde{\mathbf{t}}}_c,\bm{\rho}_e,\boldsymbol{\omega}) \label{17a}\\
\text{s.t. } \; & ||\boldsymbol{\omega}||_2^2\leq P_t, \label{17b} \\
& \tilde{\mathbf{t}}_s\in\mathcal{C}_t,\tilde{\mathbf{t}}_c\in\mathcal{C}_t,\tilde{\mathbf{r}}\in\mathcal{C}_r,\label{17c}\\
&||\mathbf{t}_{a_1}-\mathbf{t}_{a_2}||_2\geq D, 1\leq a_1\neq a_2\leq N,\label{17d}\\
&||\mathbf{r}_{b_1}-\mathbf{r}_{b_2}||_2\geq D, 1\leq b_1\neq b_2\leq M,\label{17e}\\
&\text{CRB}_{\alpha_e}(\tilde{\mathbf{t}}_s,\tilde{\mathbf{r}})\leq \eta, \text{CRB}_{\beta_e}(\tilde{\mathbf{t}}_s,\tilde{\mathbf{r}})\leq \eta.\label{17f}
\end{align}
\end{subequations}
Here, \eqref{17b} is the maximum transmit power constraint for communication. Constraint \eqref{17c} confines the MAs to their designated regions. Constraints \eqref{17d} and \eqref{17e} enforce a minimum inter-MA distance $D$ to mitigate coupling effects. Finally, $\eta$ in \eqref{17f} denotes a predefined maximum allowable CRB threshold representing the sensing-accuracy requirement. It is treated as a fixed system design parameter rather than an optimization variable. 

This formulation highlights a key physical insight: the sensing accuracy, characterized by the CRB, is a controllable factor that directly affects the secrecy rate. Optimizing MAs’ positions to reduce the CRB-based uncertainty region facilitates more effective robust beamforming design, which is a new capability that is fundamentally absent in FPA systems. 

Problem \eqref{17} is intractable due to the complex max-min objective function, the non-convex constraints in \eqref{17d}-\eqref{17f}, and the strong coupling among the optimization variables. To address this challenge, we decompose the problem into two sequential subproblems.

First, we address the optimization of MA positions for the eavesdropper sensing stage. The objective is to minimize the CRBs, which implicitly improves the worst-case secrecy rate by reducing estimation uncertainty. This leads to the following CRB minimization problem:
\begin{subequations}  \label{18}
\begin{align}
\max_{{\tilde{\mathbf{t}}}_s,\tilde{\mathbf{r}},\bar{\eta}}&\text{ } \bar{\eta} \label{18a}\\
\text{s.t. }& \tilde{\mathbf{t}}_s\in\mathcal{C}_t,\tilde{\mathbf{r}}\in\mathcal{C}_r,\label{18b}\\
& v(\mathbf{x}_t^s)+v(\mathbf{x}_r^s)-\frac{(c(\mathbf{x}_t^s,\mathbf{y}_t^s)+c(\mathbf{x}_r^s,\mathbf{y}_r^s))^2}{v(\mathbf{y}_t^s)+v(\mathbf{y}_r^s)}\geq \bar{\eta},\label{18c}\\
& v(\mathbf{y}_t^s)+v(\mathbf{y}_r^s)-\frac{(c(\mathbf{x}_t^s,\mathbf{y}_t^s)+c(\mathbf{x}_r^s,\mathbf{y}_r^s))^2}{v(\mathbf{x}_t^s)+v(\mathbf{x}_r^s)}\geq \bar{\eta},\label{18d}\\
&\eqref{17d}, \eqref{17e},
\end{align}
\end{subequations}
where $\bar{\eta}=\frac{G}{\eta}$, and \eqref{18c} and \eqref{18d} are simplified from \eqref{16}.

Second, after obtaining the optimized sensing positions and the resulting AoD uncertainty region, we solve the beamforming and MA position optimization for the secure communication stage. The goal is to jointly optimize ${\tilde{\mathbf{t}}}_c$ and $\boldsymbol{\omega}$ to maximize the worst-case secrecy rate:
\begin{subequations}  \label{19}
\begin{align}
\max_{\tilde{\mathbf{t}}_c,\boldsymbol{\omega}}&\min_{\hat{\bm{\rho}}_e\in\Xi_1,\bm{\rho}_e\in\Xi_2}\text{ } R_{\text{sec}}({\tilde{\mathbf{t}}}_c,\bm{\rho}_e,\boldsymbol{\omega}) \label{19a}\\
\text{s.t. }& \tilde{\mathbf{t}}_c \in \mathcal{C}_t, \label{19b}\\
&\eqref{17b}, \eqref{17d}.
\end{align}
\end{subequations}
The solutions to these two subproblems are detailed in the subsequent sections. The two-stage scheme further demonstrates the functional versatility of antenna mobility, where MAs’ positions are adaptively reconfigured to prioritize sensing accuracy in the first stage and communication security in the second stage.

\section{Optimization of MA Positions for Eavesdropper Sensing}
The CRB derived in Section III-A explicitly reveals how the AoD estimation accuracy depends on the MAs’ positions. This indicates that the sensing performance is not a fixed system parameter, but rather a dynamic metric that can be actively enhanced through MA position optimization. Therefore, by carefully adjusting the transmit and receive MAs’ positions $({\tilde{\mathbf{t}}}_s$ and $\tilde{\mathbf{r}})$, the AoD estimation uncertainty can be reduced. This step thus leads to highly accurate eavesdropper location information, which is critical for the subsequent secure communication stage. In this section, we develop a solution for problem \eqref{18} to determine the optimal ${\tilde{\mathbf{t}}}_s$ and $\tilde{\mathbf{r}}$. The problem is challenging due to the non-convex constraints \eqref{17d}, \eqref{17e}, \eqref{18c}, and \eqref{18d}. To find a high-quality, locally optimal solution, we propose an alternating optimization (AO) algorithm. Specifically, we decompose problem \eqref{18} into four subproblems and iteratively optimize the coordinate vectors $\mathbf{x}_t^s$, $\mathbf{y}_t^s$, $\mathbf{x}_r^s$, and $\mathbf{y}_r^s$ in an alternating manner until convergence.

\subsection{Optimization of $\mathbf{x}_t^s$ with Given $\mathbf{y}_t^s$, $\mathbf{x}_r^s$, and $\mathbf{y}_r^s$}
With $\mathbf{y}_t^s$, $\mathbf{x}_r^s$, and $\mathbf{y}_r^s$ fixed, the subproblem for optimizing $\mathbf{x}_t^s$ is formulated as
\begin{subequations}  \label{20}
\begin{align}
\max_{\mathbf{x}_t^s,\bar{\eta}}&\text{ } \bar{\eta}  \label{20a}\\
\text{s.t. }&\eqref{17d}, \eqref{18b}, \eqref{18c}, \eqref{18d}.
\end{align}
\end{subequations}
To handle the non-convexity, we employ the successive convex approximation (SCA) technique. First, we address the inter-MA distance constraint \eqref{17d}. The convex function $||\mathbf{t}_{a_1}-\mathbf{t}_{a_2}||_2$ can be globally lower-bounded by its first-order Taylor approximation. At a feasible point $(\mathbf{t}_{a_1}^{(k)}-\mathbf{t}_{a_2}^{(k)})$ from the $k$-th SCA iteration, where ${\mathbf{t}_{a_1}^{(k)}=[x_{a_1}^{t\;(k)},y_{a_1}^{t}]^T}$, we have
\begin{align}\label{21}
&||\mathbf{t}_{a_1}-\mathbf{t}_{a_2}||_2\nonumber\\
\geq&||\mathbf{t}_{a_1}^{(k)}-\mathbf{t}_{a_2}^{(k)}||_2\nonumber\\
\quad&+(\nabla||\mathbf{t}_{a_1}^{(k)}-\mathbf{t}_{a_2}^{(k)}||_2)^T((\mathbf{t}_{a_1}-\mathbf{t}_{a_2})-(\mathbf{t}_{a_1}^{(k)}-\mathbf{t}_{a_2}^{(k)}))\nonumber\\
=&\frac{(\mathbf{t}_{a_1}^{(k)}-\mathbf{t}_{a_2}^{(k)})^T(\mathbf{t}_{a_1}-\mathbf{t}_{a_2})}{||\mathbf{t}_{a_1}^{(k)}-\mathbf{t}_{a_2}^{(k)}||_2}.
\end{align}
This transforms the non-convex constraint \eqref{17d} into the following linear constraint:
\begin{align}\label{22}
\frac{(\mathbf{t}_{a_1}^{(k)}-\mathbf{t}_{a_2}^{(k)})^T(\mathbf{t}_{a_1}-\mathbf{t}_{a_2})}{||\mathbf{t}_{a_1}^{(k)}-\mathbf{t}_{a_2}^{(k)}||_2}\geq D, \; 1\leq a_1\neq a_2\leq N.
\end{align}
Next, we convexify the non-convex constraints \eqref{18c} and \eqref{18d}. We express the variance and covariance terms in standard quadratic form:  
\begin{align}
v(\mathbf{x}_t^s)&=(\mathbf{x}_t^s)^T\mathbf{P}_1\mathbf{x}_t^s,\quad\quad\,\, v(\mathbf{y}_t^s)=(\mathbf{y}_t^s)^T\mathbf{P}_1\mathbf{y}_t^s,\label{23}\\
v(\mathbf{x}_r^s)&=(\mathbf{x}_r^s)^T\mathbf{P}_2\mathbf{x}_r^s,\quad\quad\,\, v(\mathbf{y}_r^s)=(\mathbf{y}_r^s)^T\mathbf{P}_2\mathbf{y}_r^s,\label{24}\\
c(\mathbf{x}_t^s,\mathbf{y}_t^s)&=(\mathbf{x}_t^s)^T\mathbf{P}_1\mathbf{y}_t^s,\quad c(\mathbf{x}_r^s,\mathbf{y}_r^s)=(\mathbf{x}_r^s)^T\mathbf{P}_2\mathbf{y}_r^s,\label{25}
\end{align}
where $\mathbf{P}_1\triangleq\frac{1}{N}\mathbf{I}_N-\frac{1}{N^2}\mathbf{1}_N\mathbf{1}_N^T$ and $\mathbf{P}_2\triangleq\frac{1}{M}\mathbf{I}_M-\frac{1}{M^2}\mathbf{1}_M\mathbf{1}_M^T$ are positive semi-definites (PSDs). Therefore, \eqref{18c} and \eqref{18d} can be expressed as
\begin{align}
&((\mathbf{x}_t^s)^T\mathbf{P}_1\mathbf{x}_t^s+(\mathbf{x}_r^s)^T\mathbf{P}_2\mathbf{x}_r^s)((\mathbf{y}_t^s)^T\mathbf{P}_1\mathbf{y}_t^s+(\mathbf{y}_r^s)^T\mathbf{P}_2\mathbf{y}_r^s)-\nonumber\\
&((\mathbf{x}_t^s)^T\mathbf{P}_1\mathbf{y}_t^s+(\mathbf{x}_r^s)^T\mathbf{P}_2\mathbf{y}_r^s)^2\geq \bar{\eta}((\mathbf{y}_t^s)^T\mathbf{P}_1\mathbf{y}_t^s+(\mathbf{y}_r^s)^T\mathbf{P}_2\mathbf{y}_r^s),\label{26}\\
&((\mathbf{x}_t^s)^T\mathbf{P}_1\mathbf{x}_t^s+(\mathbf{x}_r^s)^T\mathbf{P}_2\mathbf{x}_r^s)((\mathbf{y}_t^s)^T\mathbf{P}_1\mathbf{y}_t^s+(\mathbf{y}_r^s)^T\mathbf{P}_2\mathbf{y}_r^s)-\nonumber\\
&((\mathbf{x}_t^s)^T\mathbf{P}_1\mathbf{y}_t^s+(\mathbf{x}_r^s)^T\mathbf{P}_2\mathbf{y}_r^s)^2\geq \bar{\eta}((\mathbf{x}_t^s)^T\mathbf{P}_1\mathbf{x}_t^s+(\mathbf{x}_r^s)^T\mathbf{P}_2\mathbf{x}_r^s).\label{27}
\end{align}
Denote $\mathbf{x}_t^{s\;(k)}$ as the solution of $\mathbf{x}_t^s$ in the $k$-th SCA iteration. Since the quadratic form $(\mathbf{x}_t^s)^T\mathbf{P}_1\mathbf{x}_t^s$ is convex, the following inequality holds via its first-order Taylor expansion at $\mathbf{x}_t^{s\;(k)}$ as
\begin{align}
(\mathbf{x}_t^s)^T\mathbf{P}_1\mathbf{x}_t^s&\geq(\mathbf{x}_t^{s\;(k)})^T\mathbf{P}_1\mathbf{x}_t^{s\;(k)}+2(\mathbf{x}_t^{s\;(k)})^T\mathbf{P}_1(\mathbf{x}_t^s-\mathbf{x}_t^{s\;(k)})\nonumber\\
&=2(\mathbf{x}_t^{s\;(k)})^T\mathbf{P}_1\mathbf{x}_t^s-(\mathbf{x}_t^{s\;(k)})^T\mathbf{P}_1\mathbf{x}_t^{s\;(k)}.\label{28}
\end{align}
Then, \eqref{26} and \eqref{27} can be written as 
\begin{align}
&(2(\mathbf{x}_t^{s\;(k)})^T\mathbf{P}_1\mathbf{x}_t^s-(\mathbf{x}_t^{s\;(k)})^T\mathbf{P}_1\mathbf{x}_t^{s\;(k)}+(\mathbf{x}_r^s)^T\mathbf{P}_2\mathbf{x}_r^s-\bar{\eta})\nonumber\\
&\times((\mathbf{y}_t^s)^T\mathbf{P}_1\mathbf{y}_t^s+(\mathbf{y}_r^s)^T\mathbf{P}_2\mathbf{y}_r^s)-((\mathbf{x}_t^s)^T\mathbf{P}_1\mathbf{y}_t^s)^2\nonumber\\
&-((\mathbf{x}_r^s)^T\mathbf{P}_2\mathbf{y}_r^s)^2-2(\mathbf{x}_t^s)^T\mathbf{P}_1\mathbf{y}_t^s(\mathbf{x}_r^s)^T\mathbf{P}_2\mathbf{y}_r^s\geq 0,\label{29}\\
&(2(\mathbf{x}_t^{s\;(k)})^T\mathbf{P}_1\mathbf{x}_t^s-(\mathbf{x}_t^{s\;(k)})^T\mathbf{P}_1\mathbf{x}_t^{s\;(k)}+(\mathbf{x}_r^s)^T\mathbf{P}_2\mathbf{x}_r^s)\nonumber\\
&\times((\mathbf{y}_t^s)^T\mathbf{P}_1\mathbf{y}_t^s+(\mathbf{y}_r^s)^T\mathbf{P}_2\mathbf{y}_r^s-\bar{\eta})-((\mathbf{x}_t^s)^T\mathbf{P}_1\mathbf{y}_t^s)^2\nonumber\\
&-((\mathbf{x}_r^s)^T\mathbf{P}_2\mathbf{y}_r^s)^2-2(\mathbf{x}_t^s)^T\mathbf{P}_1\mathbf{y}_t^s(\mathbf{x}_r^s)^T\mathbf{P}_2\mathbf{y}_r^s\geq 0.\label{30}
\end{align}
These constraints are convex quadratic w.r.t. $\mathbf{x}_t^s$ and linear w.r.t. $\bar{\eta}$. Thus, for the $(k+1)$-th SCA iteration, we solve the following convex optimization problem:
\begin{subequations} \label{31}
\begin{align}
\max_{\mathbf{x}_t^s,\bar{\eta}}&\text{ } \bar{\eta}  \label{31a}\\
\text{s.t. }& \eqref{18b}, \eqref{22}, \eqref{29}, \eqref{30},
\end{align}
\end{subequations}
which can be efficiently solved using standard convex programming solvers such as CVX \cite{RF2-8}.

\subsection{Optimization of $\mathbf{y}_t^s$ with Given $\mathbf{x}_t^s$, $\mathbf{x}_r^s$, and $\mathbf{y}_r^s$}
Next, we optimize $\mathbf{y}_t^s$ in problem \eqref{18} while keeping $\mathbf{x}_t^s$, $\mathbf{x}_r^s$, and $\mathbf{y}_r^s$ fixed. Following a procedure analogous to that in Section IV-A, we use SCA to convexify the problem. For brevity, we omit the detailed derivations. In the $l$-th SCA iteration, constraint \eqref{17d} is rewritten as
\begin{align}\label{33}
\frac{(\mathbf{t}_{a_1}^{(l)}-\mathbf{t}_{a_2}^{(l)})^T(\mathbf{t}_{a_1}-\mathbf{t}_{a_2})}{||\mathbf{t}_{a_1}^{(l)}-\mathbf{t}_{a_2}^{(l)}||_2}\geq D, \; 1\leq a_1\neq a_2\leq N,
\end{align}
where ${\mathbf{t}_{a_1}^{(l)}=[x_{a_1}^{t},y_{a_1}^{t\;(l)}]^T}$ and ${\mathbf{t}_{a_2}^{(l)}=[x_{a_2}^{t}},y_{a_2}^{t\;(l)}]^T$. 
Besides, \eqref{18c} and \eqref{18d} are transformed to 
\begin{align}
&(2(\mathbf{y}_t^{s\;(l)})^T\mathbf{P}_1\mathbf{y}_t^s-(\mathbf{y}_t^{s\;(l)})^T\mathbf{P}_1\mathbf{y}_t^{s\;(l)}+(\mathbf{y}_r^s)^T\mathbf{P}_2\mathbf{y}_r^s)\nonumber\\
&\times((\mathbf{x}_t^s)^T\mathbf{P}_1\mathbf{x}_t^s+(\mathbf{x}_r^s)^T\mathbf{P}_2\mathbf{x}_r^s-\bar{\eta})-((\mathbf{x}_t^s)^T\mathbf{P}_1\mathbf{y}_t^s)^2\nonumber\\
&-((\mathbf{x}_r^s)^T\mathbf{P}_2\mathbf{y}_r^s)^2-2(\mathbf{x}_t^s)^T\mathbf{P}_1\mathbf{y}_t^s(\mathbf{x}_r^s)^T\mathbf{P}_2\mathbf{y}_r^s\geq 0,\label{35}\\
&(2(\mathbf{y}_t^{s\;(l)})^T\mathbf{P}_1\mathbf{y}_t^s-(\mathbf{y}_t^{s\;(l)})^T\mathbf{P}_1\mathbf{y}_t^{s\;(l)}+(\mathbf{y}_r^s)^T\mathbf{P}_2\mathbf{y}_r^s-\bar{\eta})\nonumber\\
&\times((\mathbf{x}_t^s)^T\mathbf{P}_1\mathbf{x}_t^s+(\mathbf{x}_r^s)^T\mathbf{P}_2\mathbf{x}_r^s)-((\mathbf{x}_t^s)^T\mathbf{P}_1\mathbf{y}_t^s)^2\nonumber\\
&-((\mathbf{x}_r^s)^T\mathbf{P}_2\mathbf{y}_r^s)^2-2(\mathbf{x}_t^s)^T\mathbf{P}_1\mathbf{y}_t^s(\mathbf{x}_r^s)^T\mathbf{P}_2\mathbf{y}_r^s\geq 0.\label{36}
\end{align}
Thus, the resulting convex subproblem for the $(l+1)$-th iteration can be formulated as
\begin{subequations} \label{37}
\begin{align}
\max_{\mathbf{y}_t^s,\bar{\eta}}&\text{ } \bar{\eta}  \label{37a}\\
\text{s.t. }& \eqref{18b}, \eqref{33}, \eqref{35}, \eqref{36},
\end{align}
\end{subequations}
which can be solved efficiently via CVX.

\subsection{Optimization of $\mathbf{x}_r^s$ with Given $\mathbf{x}_t^s$, $\mathbf{y}_t^s$, and $\mathbf{y}_r^s$}
Here, we optimize $\mathbf{x}_r^s$ in problem \eqref{18} with $\mathbf{x}_t^s$, $\mathbf{y}_t^s$, and $\mathbf{y}_r^s$ held constant. Again, we apply the SCA technique. At the $p$-th SCA iteration, we denote $\mathbf{r}_{b_1}^{(p)}=[x_{b_1}^{r\;(p)},y_{b_1}^{r}]^T$ and $\mathbf{r}_{b_2}^{(p)}=[x_{b_2}^{r\;(p)},y_{b_2}^{r}]^T$, and let $\mathbf{x}_r^{s\;(p)}$ denote the solution of $\mathbf{x}_r^s$. We linearize the constraint \eqref{17e} and also apply a first-order approximation to the convex quadratic term $(\mathbf{x}_r^s)^T\mathbf{P}_2\mathbf{x}_r^s$ in both \eqref{18c} and \eqref{18d}. Then, \eqref{17e}, \eqref{18c} and \eqref{18d} can be expressed as
\begin{align}
&\frac{(\mathbf{r}_{b_1}^{(p)}-\mathbf{r}_{b_2}^{(p)})^T(\mathbf{r}_{b_1}-\mathbf{r}_{b_2})}{||\mathbf{r}_{b_1}^{(p)}-\mathbf{r}_{b_2}^{(p)}||_2}\geq D, \; 1\leq b_1\neq b_2\leq M,\label{39}\\
&(2(\mathbf{x}_r^{s\;(p)})^T\mathbf{P}_2\mathbf{x}_r^s-(\mathbf{x}_r^{s\;(p)})^T\mathbf{P}_2\mathbf{x}_r^{s\;(p)}+(\mathbf{x}_t^s)^T\mathbf{P}_1\mathbf{x}_t^s-\bar{\eta})\nonumber\\
&\times((\mathbf{y}_t^s)^T\mathbf{P}_1\mathbf{y}_t^s+(\mathbf{y}_r^s)^T\mathbf{P}_2\mathbf{y}_r^s)-((\mathbf{x}_t^s)^T\mathbf{P}_1\mathbf{y}_t^s)^2\nonumber\\
&-((\mathbf{x}_r^s)^T\mathbf{P}_2\mathbf{y}_r^s)^2-2(\mathbf{x}_t^s)^T\mathbf{P}_1\mathbf{y}_t^s(\mathbf{x}_r^s)^T\mathbf{P}_2\mathbf{y}_r^s\geq 0,\label{41}\\
&(2(\mathbf{x}_r^{s\;(p)})^T\mathbf{P}_2\mathbf{x}_r^s-(\mathbf{x}_r^{s\;(p)})^T\mathbf{P}_2\mathbf{x}_r^{s\;(p)}+(\mathbf{x}_t^s)^T\mathbf{P}_1\mathbf{x}_t^s)\nonumber\\
&\times((\mathbf{y}_t^s)^T\mathbf{P}_1\mathbf{y}_t^s+(\mathbf{y}_r^s)^T\mathbf{P}_2\mathbf{y}_r^s-\bar{\eta})-((\mathbf{x}_t^s)^T\mathbf{P}_1\mathbf{y}_t^s)^2\nonumber\\
&-((\mathbf{x}_r^s)^T\mathbf{P}_2\mathbf{y}_r^s)^2-2(\mathbf{x}_t^s)^T\mathbf{P}_1\mathbf{y}_t^s(\mathbf{x}_r^s)^T\mathbf{P}_2\mathbf{y}_r^s\geq 0.\label{42}
\end{align}
Thus, the resulting problem in the $(p+1)$-th SCA iteration is 
\begin{subequations} \label{43}
\begin{align}
\max_{\mathbf{x}_r^s,\bar{\eta}}&\text{ } \bar{\eta}  \label{43a}\\
\text{s.t. }& \eqref{18b}, \eqref{39}, \eqref{41}, \eqref{42},
\end{align}
\end{subequations}
which can be solved by CVX efficiently.

\subsection{Optimization of $\mathbf{y}_r^s$ with Given $\mathbf{x}_t^s$, $\mathbf{y}_t^s$, and $\mathbf{x}_r^s$}
Finally, we optimize $\mathbf{y}_r^s$ in problem \eqref{18} while fixing the other variables. The structure of this subproblem is identical to that for optimizing $\mathbf{x}_r^s$. Therefore, we apply the same SCA-based approach. At the $q$-th SCA iteration, we denote $\mathbf{r}_{b_1}^{(q)}=[x_{b_1}^{r},y_{b_1}^{r\;(q)}]^T$ and $\mathbf{r}_{b_2}^{(q)}=[x_{b_2}^{r},y_{b_2}^{r\;(q)}]^T$, and let $\mathbf{y}_r^{(q)}$ denote the solution of $\mathbf{y}_r^s$. By reformulating \eqref{17e}, \eqref{18c}, and \eqref{18d} as
\begin{align}
&\frac{(\mathbf{r}_{b_1}^{(q)}-\mathbf{r}_{b_2}^{(q)})^T(\mathbf{r}_{b_1}-\mathbf{r}_{b_2})}{||\mathbf{r}_{b_1}^{(q)}-\mathbf{r}_{b_2}^{(q)}||_2}\geq D, \; 1\leq b_1\neq b_2\leq M,\label{45}\\
&(2(\mathbf{y}_r^{s\;(q)})^T\mathbf{P}_2\mathbf{y}_r^s-(\mathbf{y}_r^{s\;(q)})^T\mathbf{P}_2\mathbf{y}_r^{s\;(q)}+(\mathbf{y}_t^s)^T\mathbf{P}_1\mathbf{y}_t^s)\nonumber\\
&\times((\mathbf{x}_t^s)^T\mathbf{P}_1\mathbf{x}_t^s+(\mathbf{x}_r^s)^T\mathbf{P}_2\mathbf{x}_r^s-\bar{\eta})-((\mathbf{x}_t^s)^T\mathbf{P}_1\mathbf{y}_t^s)^2\nonumber\\
&-((\mathbf{x}_r^s)^T\mathbf{P}_2\mathbf{y}_r^s)^2-2(\mathbf{x}_t^s)^T\mathbf{P}_1\mathbf{y}_t^s(\mathbf{x}_r^s)^T\mathbf{P}_2\mathbf{y}_r^s\geq 0,\label{47}\\
&(2(\mathbf{y}_r^{s\;(q)})^T\mathbf{P}_2\mathbf{y}_r^s-(\mathbf{y}_r^{s\;(q)})^T\mathbf{P}_2\mathbf{y}_r^{s\;(q)}+(\mathbf{y}_t^s)^T\mathbf{P}_1\mathbf{y}_t^s-\bar{\eta})\nonumber\\
&\times((\mathbf{x}_t^s)^T\mathbf{P}_1\mathbf{x}_t^s+(\mathbf{x}_r^s)^T\mathbf{P}_2\mathbf{x}_r^s)-((\mathbf{x}_t^s)^T\mathbf{P}_1\mathbf{y}_t^s)^2\nonumber\\
&-((\mathbf{x}_r^s)^T\mathbf{P}_2\mathbf{y}_r^s)^2-2(\mathbf{x}_t^s)^T\mathbf{P}_1\mathbf{y}_t^s(\mathbf{x}_r^s)^T\mathbf{P}_2\mathbf{y}_r^s\geq 0,\label{48}
\end{align}
respectively, we solve the below problem in the $(q+1)$-th SCA iteration using CVX to update $\mathbf{y}_r^s$.
\begin{subequations} \label{49}
\begin{align}
\max_{\mathbf{y}_r^s,\bar{\eta}}&\text{ } \bar{\eta}  \label{49a}\\
\text{s.t. }& \eqref{18b}, \eqref{45}, \eqref{47}, \eqref{48}.
\end{align}
\end{subequations}

\subsection{Overall Algorithm for Problem \eqref{18}}
Based on the solutions to these four subproblems, we present the overall AO framework for problem \eqref{18} in Algorithm 1. To mitigate the sensitivity of the AO algorithm to the initial positions $\mathbf{x}_t^{s\;(0)}, {\ \mathbf{y}}_t^{s\;(0)}, {\ \mathbf{x}}_r^{s\;(0)},$ and ${\mathbf{y}}_r^{s\;(0)}$, we employ a multiple random initialization strategy for the MAs. Specifically, $M_r$ sets of initial positions for the transmit and receive MAs are randomly generated within their respective feasible movement regions. Algorithm 1 is executed for each initialization, and the solution yielding the maximum objective value $\bar{\eta}$ is selected as the final solution. In Algorithm 1, the predefined convergence thresholds are denoted by $\delta_1$ for the objective value in \eqref{18a}, and $\delta_2$ for those in \eqref{31a}, \eqref{37a}, \eqref{43a}, and \eqref{49a}, respectively. The four subproblems \eqref{31}, \eqref{37}, \eqref{43}, and \eqref{49} are solved iteratively until the increase in \eqref{18a} falls below $\delta_1$. Moreover, each subproblem is solved iteratively until the increase in its own objective value falls below $\delta_2$.

The computational complexity of Algorithm 1 is analyzed below. Using the interior-point method with solution accuracy $\kappa$, the computational complexities of solving problems \eqref{31} and \eqref{37} are $O(N^{3.5} \ln(1/\kappa))$, and those of solving problems \eqref{43} and \eqref{49} are $O(M^{3.5} \ln(1/\kappa))$ \cite{RF2-9}. We define $U$, $K$, $L$, $P$, and $Q$ as the maximum iteration numbers of steps 2-26, steps 3-6, steps 8-11, steps 13-16, and steps 18-21, respectively. Thus, the total computational complexity is $\mathcal{O}(UM_r ( N^{3.5} \ln(1/\kappa)(K + L) + M^{3.5} \ln(1/\kappa)(P + Q)))$. 

Furthermore, although Algorithm 1 involves nested AO and SCA procedures, it converges within a moderate number of iterations, as shown in Section VI-A. Moreover, since the MA position optimization depends mainly on slowly varying angular parameters (e.g., AoDs), it can be performed offline or updated infrequently over multiple transmission intervals, which alleviates the real-time computational burden at the BS.

\begin{algorithm}[!t]
\caption{MA Position Optimization for Eavesdropper Sensing}
\label{alg1}
\textbf{Input:} $N$, $M$, $D$, $C$, $\mathcal{C}_t$, $\mathcal{C}_r$, $\mathbf{x}_{t}^{s\;(0)}$, $\mathbf{y}_{t}^{s\;(0)}$, $\mathbf{x}_{r}^{s\;(0)}$, $\mathbf{y}_{r}^{s\;(0)}$, $\mathbf{P}_1$, $\mathbf{P}_2$, $\delta_1$, $\delta_2$.\\
\textbf{Output:} $\mathbf{x}_t^s$, $\mathbf{y}_t^s$, $\mathbf{x}_r^s$, $\mathbf{y}_r^s$.
\begin{algorithmic}[1]
    \State \parbox[t]{0.4\textwidth}{Initialization: $k\leftarrow0$, $l\leftarrow0$, $p\leftarrow0$, $q\leftarrow0$.}
    \While{Increase of $\bar{\eta}$ in \eqref{18a} is above $\delta_1$}
        \While{Increase of $\bar{\eta}$ in \eqref{31a} is above $\delta_2$}
        \State Obtain $\mathbf{x}_t^{s\;(k+1)}$ by solving problem \eqref{31}.
        \State $k\leftarrow k+1$.
        \EndWhile
        \State $\mathbf{x}_t^s\leftarrow \mathbf{x}_{t}^{s\;(k)}$, $k\leftarrow 0$.
        \While{Increase of $\bar{\eta}$ in \eqref{37a} is above $\delta_2$}
        \State Obtain $\mathbf{y}_t^{s\;(l+1)}$ by solving problem \eqref{37}.
        \State $l\leftarrow l+1$.
        \EndWhile
        \State $\mathbf{y}_t^s\leftarrow \mathbf{y}_{t}^{s\;(l)}$, $l\leftarrow0$.
        \While{Increase of $\bar{\eta}$ in \eqref{43a} is above $\delta_2$}
        \State Obtain $\mathbf{x}_r^{s\;(p+1)}$ by solving problem \eqref{43}.
        \State $p\leftarrow p+1$.
        \EndWhile
        \State $\mathbf{x}_r^s\leftarrow \mathbf{x}_{r}^{s\;(p)}$, $p\leftarrow0$.
        \While{Increase of $\bar{\eta}$ in \eqref{49a} is above $\delta_2$}
        \State Obtain $\mathbf{y}_r^{s\;(q+1)}$ by solving problem \eqref{49}.
        \State $q\leftarrow q+1$.
        \EndWhile
        \State $\mathbf{y}_r^s\leftarrow \mathbf{y}_{r}^{s\;(q)}$, $q\leftarrow0$.
        \State $\bar{\eta}_1\leftarrow v(\mathbf{x}_t^s)+v(\mathbf{x}_r^s)-\frac{(c(\mathbf{x}_t^s,\mathbf{y}_t^s)+c(\mathbf{x}_r^s,\mathbf{y}_r^s))^2}{v(\mathbf{y}_t^s)+v(\mathbf{y}_r^s)}$.
        \State $\bar{\eta}_2\leftarrow v(\mathbf{y}_t^s)+v(\mathbf{y}_r^s)-\frac{(c(\mathbf{x}_t^s,\mathbf{y}_t^s)+c(\mathbf{x}_r^s,\mathbf{y}_r^s))^2}{v(\mathbf{x}_t^s)+v(\mathbf{x}_r^s)}$.
        \State $\bar{\eta}\leftarrow \min\{\bar{\eta}_1,\bar{\eta}_2\}$.
    \EndWhile 
\end{algorithmic}
\end{algorithm}

\section{Optimization of Beamforming and MA Positions for Secure Communication}
Since the AoD uncertainty region affects the spatial characteristics of the worst-case eavesdropper channel, jointly optimizing the beamforming vector $\boldsymbol{\omega}$ and the transmit MAs' positions $\tilde{\mathbf{t}}_c$ provides additional spatial DoFs to improve the secrecy rate under AoD uncertainty. In this section, we address the robust design for the secure communication stage, aiming to maximize the worst-case secrecy rate given the AoD uncertainty established in the sensing stage. The core task is to solve problem \eqref{19} by jointly optimizing $\boldsymbol{\omega}$ and $\tilde{\mathbf{t}}_c$. The key difference between our proposed method and conventional robust beamforming lies in the fact that the estimation uncertainty in our system is not {\it a priori} known. Instead, it is  determined by the sensing accuracy, which explicitly depends on the optimized MAs’ positions, as characterized by the CRB in the eavesdropper sensing stage.

Since an analytical characterization of the worst-case secrecy rate over all possible estimates $\hat{\bm{\rho}}_e \in \Xi_1$ is intractable, we approximate it by considering the estimate that yields the lowest secrecy rate from a set of Monte Carlo trials. For a given worst-case estimate $\hat{\bm{\rho}}_e$ and its corresponding uncertainty region $\bm{\rho}_e\in\Xi_2$, problem \eqref{19} simplifies to
\begin{subequations} \label{50}
\begin{align}
\max_{\tilde{\mathbf{t}}_c,\boldsymbol{\omega}}&\min_{\bm{\rho}_e\in\Xi_2}\text{ } R_{\text{sec}}({\tilde{\mathbf{t}}}_c,\bm{\rho}_e,\boldsymbol{\omega}) \label{50a}\\
\text{s.t. }&\eqref{19b}, \eqref{17b}, \eqref{17d}. 
\end{align}
\end{subequations}
Note that the operator $[x]^+$ can be omitted without loss of generality, as a feasible solution $\boldsymbol{\omega}=\mathbf{0}$ always ensures a non-negative secrecy rate \cite{RF2-A}. Nevertheless, problem \eqref{50} remains challenging due to the non-concave max-min objective and the non-convex inter-MA distance constraint \eqref{17d}. To manage the strong coupling between $\tilde{\mathbf{t}}_c$ and $\boldsymbol{\omega}$, we decompose the problem into two subproblems and employ a backward induction approach.

\subsection{Optimization of Robust Beamforming $\boldsymbol{\omega}$ for a Given $\tilde{\mathbf{t}}_c$}
For a fixed set of transmit MA positions $\tilde{\mathbf{t}}_c$, we first optimize the beamforming vector $\boldsymbol{\omega}$ by solving the following worst-case subproblem:
\begin{subequations} \label{51}
\begin{align}
\max_{\boldsymbol{\omega}}&\min_{\bm{\rho}_e \in \Xi_2}\text{ } R_\text{sec}(\bm{\rho}_e,\boldsymbol{\omega}) \label{51a}\\
\text{s.t. }&\eqref{17b}.
\end{align}
\end{subequations}
Since the logarithm is a monotonic function, maximizing the secrecy rate is equivalent to maximizing the ratio of the legitimate receiver's SNR to the eavesdropper's SNR. Problem \eqref{51} can thus be reformulated as
\begin{subequations} \label{52}
\begin{align}
\max_{\boldsymbol{\omega}}&\min_{\mathbf{H}_e(\tilde{\mathbf{t}}_c,\bm{\rho}_e)\in\mathbf{\Gamma}_1}\text{ }\frac{\boldsymbol{\omega}^H\mathbf{H}_c(\tilde{\mathbf{t}}_c,\bm{\rho}_c)\boldsymbol{\omega}+\sigma_c^2}{\boldsymbol{\omega}^H\mathbf{H}_e(\tilde{\mathbf{t}}_c,\bm{\rho}_e)\boldsymbol{\omega}+\sigma_e^2}  \label{52a}\\
\text{s.t. }&\eqref{17b},
\end{align}
\end{subequations}
where $\mathbf{H}_c(\tilde{\mathbf{t}}_c,\bm{\rho}_c)=\mathbf{h}_c (\tilde{\mathbf{t}}_c,\bm{\rho}_c)\mathbf{h}_c^H (\tilde{\mathbf{t}}_c,\bm{\rho}_c)$, $\mathbf{H}_e(\tilde{\mathbf{t}}_c,\bm{\rho}_e)=\mathbf{h}_e (\tilde{\mathbf{t}}_c,\bm{\rho}_e)\mathbf{h}_e^H (\tilde{\mathbf{t}}_c,\bm{\rho}_e)$,
and $\mathbf{\Gamma}_1=\{\mathbf{h}_e (\tilde{\mathbf{t}}_c,\bm{\rho}_e)\mathbf{h}_e^H (\tilde{\mathbf{t}}_c,\bm{\rho}_e) | \bm{\rho}_e\in\Xi_2\}$.  To tackle this intractable max-min problem, we relax the continuous uncertainty set $\mathbf{\Gamma}_1$ to its discrete convex hull $\mathbf{\Gamma}_2$ \cite{RF2-7}, defined as
\begin{align}
\mathbf{\Gamma}_2\triangleq\left\{\sum_{f=1}^F\mu_f\mathbf{H}_{e,f}(\tilde{\mathbf{t}}_c,\bm{\rho}_e)|\sum_{f=1}^F\mu_f=1,\mu_f\geq0\right\},\label{53}
\end{align}
where $\{\mathbf{H}_{e,f}(\tilde{\mathbf{t}}_c,\bm{\rho}_e)\}_{f=1}^F$ are $F$ discrete channel samples from $\mathbf{\Gamma}_1$ and $\{\mu_f\}$ are weighting coefficients. This relaxation allows us to transform the max-min problem in \eqref{52a} into an equivalent min-max problem over the convex hull $\mathbf{\Gamma}_2$ \cite{RF2-10}, i.e., 
\begin{align}
&\max_{\bm{\omega}}\min_{\mathbf{H}_e(\tilde{\mathbf{t}}_c,\bm{\rho}_e)\in\mathbf{\Gamma}_1}\text{ }\frac{\bm{\omega}^H\mathbf{H}_c(\tilde{\mathbf{t}}_c,\bm{\rho}_c)\bm{\omega}+\sigma_c^2}{\bm{\omega}^H\mathbf{H}_e(\tilde{\mathbf{t}}_c,\bm{\rho}_e)\bm{\omega}+\sigma_e^2}\nonumber\\
=&\min_{\mathbf{H}_e(\tilde{\mathbf{t}}_c,\bm{\rho}_e)\in\mathbf{\Gamma}_2}\max_{\bm{\omega}}\text{ }\frac{\bm{\omega}^H\mathbf{H}_c(\tilde{\mathbf{t}}_c,\bm{\rho}_c)\bm{\omega}+\sigma_c^2}{\bm{\omega}^H\mathbf{H}_e(\tilde{\mathbf{t}}_c,\bm{\rho}_e)\bm{\omega}+\sigma_e^2}.\label{54}
\end{align}

Following that, by setting $\boldsymbol{\omega}=\sqrt{P}\hat{\boldsymbol{\omega}}$ and utilizing the convex hull defined in \eqref{53}, problem \eqref{52} can be restated as
\begin{subequations} \label{55}
\begin{align}
\min_{\{\mu_f\}}&\max_{\hat{\boldsymbol{\omega}},P}\text{ }\frac{P\hat{\boldsymbol{\omega}}^H\mathbf{H}_c(\tilde{\mathbf{t}}_c,\bm{\rho}_c)\hat{\boldsymbol{\omega}}+\sigma_c^2}{P\hat{\boldsymbol{\omega}}^H\sum_{f=1}^F\mu_f\mathbf{H}_{e,f}(\tilde{\mathbf{t}}_c,\bm{\rho}_e)\hat{\boldsymbol{\omega}}+\sigma_e^2}\label{55a}\\
\text{s.t. }&||\hat{\boldsymbol{\omega}}||_2^2 = 1, P\leq P_t,\label{55b}\\
&\sum_{f=1}^F\mu_f=1,\mu_f\geq0.\label{55c}
\end{align}
\end{subequations}
Then, for given $\{\mu_f\}$ and $P$, problem \eqref{55} becomes a generalized Rayleigh quotient problem:
\begin{subequations} \label{56}
\begin{align}
\max_{\hat{\boldsymbol{\omega}}}&\text{ }\frac{\hat{\boldsymbol{\omega}}^H(P\mathbf{H}_c(\tilde{\mathbf{t}}_c,\bm{\rho}_c)+\sigma_c^2\mathbf{I})\hat{\boldsymbol{\omega}}}{\hat{\boldsymbol{\omega}}^H\left(P\sum_{f=1}^F\mu_f\mathbf{H}_{e,f}(\tilde{\mathbf{t}}_c,\bm{\rho}_e)+\sigma_e^2\mathbf{I}\right)\hat{\boldsymbol{\omega}}}\label{56a}\\
\text{s.t. }&||\hat{\boldsymbol{\omega}}||_2^2 = 1.\label{56b}
\end{align}
\end{subequations}
The optimal beamforming direction $\hat{\boldsymbol{\omega}}$ and the corresponding maximum objective value $\gamma$ are
\begin{subequations} \label{57}
\begin{align}
&\hat{\boldsymbol{\omega}}=\text{eig}\bigg(P\mathbf{H}_c(\tilde{\mathbf{t}}_c,\bm{\rho}_c)+\sigma_c^2\mathbf{I},P\sum_{f=1}^F\mu_f\mathbf{H}_{e,f}(\tilde{\mathbf{t}}_c,\bm{\rho}_e)+\sigma_e^2\mathbf{I}\bigg),\label{57a}\\
&\gamma=\lambda_{\max}\bigg(P\mathbf{H}_c(\tilde{\mathbf{t}}_c,\bm{\rho}_c)+\sigma_c^2\mathbf{I},P\sum_{f=1}^F\mu_f\mathbf{H}_{e,f}(\tilde{\mathbf{t}}_c,\bm{\rho}_e)+\sigma_e^2\mathbf{I}\bigg),\label{57b}
\end{align}
\end{subequations}
respectively. Here, $\lambda_{\max}(\mathbf{A},\mathbf{B})$ and $\text{eig}(\mathbf{A},\mathbf{B})$ denote the largest generalized eigenvalue and its corresponding principal generalized eigenvector of the matrix pair $(\mathbf{A},\mathbf{B})$, respectively. According to [11, Proposition 3], \eqref{57} can be computed efficiently as 
\begin{subequations} \label{58}
\begin{align}
&\hat{\boldsymbol{\omega}}=\frac{\left(P\sum_{f=1}^F\mu_f\mathbf{H}_{e,f}(\tilde{\mathbf{t}}_c,\bm{\rho}_e)+\sigma_e^2\mathbf{I}\right)^{-1}\mathbf{h}_c}{\left\|\left(P\sum_{f=1}^F\mu_f\mathbf{H}_{e,f}(\tilde{\mathbf{t}}_c,\bm{\rho}_e)+\sigma_e^2\mathbf{I}\right)^{-1}\mathbf{h}_c\right\|_2},\label{58a}\\
&\gamma=\mathbf{h}_c^{H}\bigg(P\sum_{f=1}^F\mu_f\mathbf{H}_{e,f}(\tilde{\mathbf{t}}_c,\bm{\rho}_e)+\sigma_e^2\mathbf{I}\bigg)^{-1}\mathbf{h}_c.\label{58b}
\end{align}
\end{subequations}

Next, since the objective function in \eqref{56a} increases monotonically with $P$, the optimal transmit power is $P^*=P_t$. Then, $\{\mu_f\}$ are optimized to minimize the objective function in \eqref{55}. Applying the Cauchy–Schwarz inequality, we have 
\begin{align}
&\bigg(\hat{\boldsymbol{\omega}}^H\sum_{f=1}^F\mu_f\mathbf{H}_{e,f}(\tilde{\mathbf{t}}_c,\bm{\rho}_e)\hat{\boldsymbol{\omega}}\bigg)^2\nonumber\\
\leq&\bigg(\sum_{f=1}^F\mu_f^2\bigg) \bigg(\sum_{f=1}^F(\hat{\boldsymbol{\omega}}^H\mathbf{H}_{e,f}(\tilde{\mathbf{t}}_c,\bm{\rho}_e)\hat{\boldsymbol{\omega}})^2\bigg).\label{60}
\end{align}
Equality holds if and only if $\frac{\mu_1}{\mathbf{H}_{e,1}(\tilde{\mathbf{t}}_c,\bm{\rho}_e)}=\frac{\mu_2}{\mathbf{H}_{e,2}(\tilde{\mathbf{t}}_c,\bm{\rho}_e)}=\dots=\frac{\mu_F}{\mathbf{H}_{e,F}(\tilde{\mathbf{t}}_c,\bm{\rho}_e)}$. Subject to the constraint \eqref{55c}, the optimal $\{\mu_f\}$ that achieves the worst-case secrecy rate are
\begin{align}
\mu_f=\frac{\hat{\boldsymbol{\omega}}^H\mathbf{H}_{e,f}(\tilde{\mathbf{t}}_c,\bm{\rho}_e)\hat{\boldsymbol{\omega}}}{\sum_{f=1}^F\hat{\boldsymbol{\omega}}^H\mathbf{H}_{e,f}(\tilde{\mathbf{t}}_c,\bm{\rho}_e)\hat{\boldsymbol{\omega}}}, \; f\in\{1,\dots,F\}.\label{61}
\end{align}

The overall procedure for finding the robust beamformer is an iterative one, summarized in Algorithm 2. The variable $\mu_f\ (f\in\{1,...,F\})$ is initialized as $\mu_f^{(0)}=\frac{1}{F}$. The initial variables ${\hat{\boldsymbol{\omega}}}^{(0)}$ and $\gamma^{(0)}$ are obtained via (52) with $P=P_t$ and ${\{\mu_f^{(0)}\}}$. In Algorithm 2, $\delta_3$ denotes the convergence threshold for $\gamma$. The variables $\mu_f$,  $\hat{\boldsymbol{\omega}}$ and $\gamma$ are calculated iteratively in steps 3-5 until the increase in $\gamma$ falls below $\delta_3$. The main complexity of Algorithm 2 lies in computing the generalized eigenvector and eigenvalue, which is $\mathcal{O}(N^{3}+FN^2)$ per iteration. Let $J$ be the number of iterations; thus, the total complexity is $\mathcal{O}(J(N^{3}+FN^2))$.

\begin{algorithm}
\caption{Robust Beamforming Optimization Algorithm for Problem \eqref{51}}
\label{alg2}
\textbf{Input:} $P_t$, $\mu_f^{(0)}$, $\hat{\boldsymbol{\omega}}^{(0)}$, $\gamma^{(0)}$, $\delta_3$, $\mathbf{h}_c(\tilde{\mathbf{t}}_c,\bm{\rho}_c)$, $\mathbf{H}_{e,f}(\tilde{\mathbf{t}}_c,\bm{\rho}_e)$, $f\in\{1,\dots,F\}$.\\
\textbf{Output:} $\boldsymbol{\omega}$.
\begin{algorithmic}[1]
    \State Initialization: $j\leftarrow0$, $P\leftarrow P_t$.
    \While{Increase of $\gamma$ in \eqref{58b} is above $\delta_3$}
       \State Given $\hat{\boldsymbol{\omega}}^{(j)}$, obtain $\mu_f^{(j+1)}$ via \eqref{61}.
       \State Given $\mu_f^{(j+1)}$, obtain $\hat{\boldsymbol{\omega}}^{(j+1)}$ and $\gamma^{(j+1)}$ via \eqref{58}.
       \State $j\leftarrow j+1$.
    \EndWhile
    \State Calculate $\hat{\boldsymbol{\omega}}$ by substituting $\mu_f^{(j)}$ into \eqref{57a}, then $\boldsymbol{\omega}\leftarrow\sqrt{P_t}\hat{\boldsymbol{\omega}}$.
\end{algorithmic}
\end{algorithm}

\subsection{Optimization of MA Positions $\tilde{\mathbf{t}}_c$}
After obtaining the solution for $\boldsymbol{\omega}$ from Algorithm 2, denoted by $\boldsymbol{\omega}^*$, we then optimize the transmit MA positions $\tilde{\mathbf{t}}_c$. According to \eqref{50}, the corresponding problem can be formulated as
\begin{subequations} \label{62}
\begin{align}
\max_{\tilde{\mathbf{t}}_c}&\text{ } R_\text{sec}(\tilde{\mathbf{t}}_c,\bm{\rho}_e,\boldsymbol{\omega}^*) \label{62a}\\
\text{s.t. }&\eqref{17d}, \eqref{19b}.
\end{align}
\end{subequations}
To tackle this non-convex problem, we adopt an AO approach that iteratively optimizes the horizontal and vertical position vectors, $\mathbf{x}_t^c$ and $\mathbf{y}_t^c$, to find a locally optimal solution for $\tilde{\mathbf{t}}_c$.

\subsubsection{Optimization of $\mathbf{x}_t^c$ with Given $\mathbf{y}_t^c$}
With the vertical position vector $\mathbf{y}_t^c$ held fixed, we first optimize the horizontal position vector $\mathbf{x}_t^c$. The non-convex constraint \eqref{17d} is converted into a linear form using the SCA technique, following a procedure similar to that in \eqref{22}. This yields the following constraint for the $i$-th SCA iteration:
\begin{align}
\frac{(\mathbf{t}_{a_1}^{(i)}-\mathbf{t}_{a_2}^{(i)})^T(\mathbf{t}_{a_1}-\mathbf{t}_{a_2})}{||\mathbf{t}_{a_1}^{(i)}-\mathbf{t}_{a_2}^{(i)}||_2}\geq D, \; 1\leq a_1\neq a_2\leq N,\label{63}
\end{align}
where ${\mathbf{t}_{a_1}^{(i)}=[x_{a_1}^{t\;(i)},y_{a_1}^{t}]^T}$ and ${\mathbf{t}_{a_2}^{(i)}=[x_{a_2}^{t\;(i)}},y_{a_2}^{t}]^T$ represent the positions at the $i$-th iteration. Consequently, problem \eqref{62} is reformulated for the $(i+1)$-th iteration as
\begin{subequations} \label{64}
\begin{align}
\max_{\mathbf{x}_t^c}&\text{ } R_\text{sec}(\mathbf{x}_t^c,\bm{\rho}_e,\boldsymbol{\omega}^*) \label{64a}\\
\text{s.t. }&\eqref{19b}, \eqref{63}.
\end{align}
\end{subequations}
Problem \eqref{64} has a non-concave objective function but convex constraints, which makes it amenable to feasible direction methods \cite{RF2-11}. To proceed, constraint \eqref{63} is further transformed into the standard linear form. Let $\mathbf{Q} \in \mathbb{R}^{N(N-1)/2 \times N}$ represent the sparse matrix of current horizontal position differences, and let $\mathbf{q} \in \mathbb{R}^{N(N-1)/2 \times 1}$ be the vector of corresponding threshold values determined by the minimum distance constraint and the current antenna positions. The entries of $\mathbf{Q}$ and $\mathbf{q}$ are defined as
\begin{align}
\mathbf{Q}[\chi(a_1,a_2),a_1]&=x_{a_1}^{t\:(i)}-x_{a_2}^{t\:(i)},\label{65}\\
\mathbf{Q}[\chi(a_1,a_2),a_2]&=x_{a_2}^{t\:(i)}-x_{a_1}^{t\:(i)},\label{66}\\
\mathbf{q}[\chi(a_1,a_2)]&=D||\mathbf{t}_{a_1}^{(i)}-\mathbf{t}_{a_2}^{(i)}||_2-(y_{a_1}^t-y_{a_2}^t)^2, \label{67}
\end{align}
where $\chi(a_1,a_2)\triangleq(2N-a_1)(a_1-1)/2+a_2-a_1$ is an index mapping function that converts the two-dimensional antenna pair index $(a_1,a_2)$ into a one-dimensional linear index. The condition $a_1<a_2$ is imposed to ensure each antenna pair is uniquely represented. Thus, constraint \eqref{63} can be written as
\begin{align}
\mathbf{Q}\mathbf{x}_t^c\geq \mathbf{q}.\label{68}
\end{align}

Next, we compute the gradient of the objective function with respect to $\mathbf{x}_t^c$, denoted by $\triangledown_{\mathbf{x}} R_\text{sec}(\mathbf{x}_t^{c\;(i)},\bm{\rho}_e,\boldsymbol{\omega}^*)$, at the current point $\mathbf{x}_t^{c\;(i)}$. The $n$-th element of this gradient can be computed as
\begin{align}
&\triangledown_{\mathbf{x}} R_\text{sec}(\mathbf{x}_t^{c\;(i)},\bm{\rho}_e,\boldsymbol{\omega}^*)[n]\nonumber\\
=&\lim_{\delta\to0}\frac{ R_\text{sec}(\mathbf{x}_t^{c\;(i)}+\delta \mathbf{e}_n,\bm{\rho}_e,\boldsymbol{\omega}^*)- R_\text{sec}(\mathbf{x}_t^{c\;(i)},\bm{\rho}_e,\boldsymbol{\omega}^*)}{\delta},\label{69}
\end{align}
where $\mathbf{e}_n$ denotes a basis vector in an $N$-dimensional space whose $n$-th element is 1 and all other elements are 0. With the gradient, we find an ascent direction $\mathbf{d}$ by solving the following linear program (LP):
\begin{subequations} \label{70}
\begin{align}
\max_{\mathbf{d}}&\text{ }\mathbf{d}^T \triangledown_{\mathbf{x}} R_\text{sec}(\mathbf{x}_t^{c\;(i)},\bm{\rho}_e,\boldsymbol{\omega}^*) \label{70a}\\
\text{s.t. }&\mathbf{t}_n(\mathbf{d})\in \mathcal{C}_t,\label{70b}\\
&\eqref{68},
\end{align}
\end{subequations}
where $\mathbf{t}_n(\mathbf{d})\triangleq[\mathbf{d}[n],y^t_n]^T$, with $\mathbf{d}$ denoting the adjustment direction of the horizontal position vector. Constraint \eqref{70b} ensures that the candidate position of each antenna along direction $\mathbf{d}$ remains within the movement region $\mathcal{C}_t$. Problem \eqref{70} is a linear optimization problem since the objective function \eqref{70a} and constraints \eqref{70b}, \eqref{68} are linear. This LP can be solved efficiently using standard optimization solvers.

After obtaining the ascent direction $\mathbf{d}$, the optimal step size $s$ is determined by performing a one-dimensional exhaustive search over the interval $[0, 1]$ to solve:
\begin{align}
s = \arg\max_{\hat{s} \in [0,1]} R_\text{sec}\left(\mathbf{x}_t^{c\;(i)} + \hat{s}(\mathbf{d} - \mathbf{x}_t^{c\;(i)}), \bm{\rho}_e, \boldsymbol{\omega}^*\right). \label{71}
\end{align}

Finally, the horizontal positions are updated using
\begin{align}
\mathbf{x}_t^{c\;(i+1)}=\mathbf{x}_t^{c\;(i)}+s(\mathbf{d}-\mathbf{x}_t^{c\;(i)}). \label{72}
\end{align}

This iterative procedure for solving problem \eqref{64} is summarized in Algorithm 3. The variable $\mathbf{x}_t^c$ is initialized as $\mathbf{x}_t^{c\ (0)}$ = $\mathbf{x}_t^{s}$. $\delta_4$ denotes the convergence threshold for $R_\text{sec}(\mathbf{x}_t^{c\ (i)},\bm{\rho}_e,\boldsymbol{\omega}^*)$. The variable $\mathbf{x}_t^{c}$ is updated iteratively by using the feasible direction methods in steps 3-7 until the increase in  $R_\text{sec}(\mathbf{x}_t^{c\ (i)},\bm{\rho}_e,\boldsymbol{\omega}^*)$ falls below $\delta_4$. The computational complexity of Algorithm 3 is dominated by the gradient calculation, the LP solver, and the line search, resulting in a total complexity of $\mathcal{O}(L_x((M_s+N)J(N^{3}+FN^2)+N^{3.5} \ln(1/\kappa)))$, where $L_x$ is the number of iterations, $M_s$ is the line search resolution, and $\kappa$ is the solution accuracy.

\begin{algorithm}[!t]
\caption{MA Position Optimization Algorithm for Problem \eqref{64}}
\label{alg3}
\textbf{Input:} $N$, $D$, $\mathcal{C}_t$, $\delta_4$, $\mathbf{x}_t^{c\;(0)}$, $\mathbf{y}_t^c$. \\
\textbf{Output:} $\mathbf{x}_t^c$.
\begin{algorithmic}[1]
    \State Initialization: $i\leftarrow0$.
    \While{Increase of $R_\text{sec}(\mathbf{x}_t^{c\;(i)},\bm{\rho}_e,\boldsymbol{\omega}^*)$ is above $\delta_4$}
       \State Compute the gradient $\triangledown_{\mathbf{x}} R_\text{sec}(\mathbf{x}_t^{c\;(i)},\bm{\rho}_e,\boldsymbol{\omega}^*)$ by \eqref{69}.
       \State Find $\mathbf{d}$ by solving problem \eqref{70}.
       \State Search $s$ by \eqref{71}.
       \State Obtain $\mathbf{x}_t^{c\;(i+1)}$ by \eqref{72}.
       \State $i\leftarrow i+1$.
    \EndWhile
\end{algorithmic}
\end{algorithm}

\subsubsection{Optimization of $\mathbf{y}_t^c$ with Given $\mathbf{x}_t^c$}
We now address the optimization of the vertical position vector $\mathbf{y}_t^c$ while holding $\mathbf{x}_t^c$ fixed. The corresponding subproblem is
\begin{subequations} \label{73}
\begin{align}
\max_{\mathbf{y}_t^c}&\text{ }R_\text{sec}(\mathbf{y}_t^c, \bm{\rho}_e, \boldsymbol{\omega}^*) \label{73a} \\
\text{s.t. }& \eqref{19b}, \eqref{63}.
\end{align}
\end{subequations}
Since problem \eqref{73} shares an identical structure with problem \eqref{64}, Algorithm 3 can be directly applied to determine the optimal vertical position vector $\mathbf{y}_t^c$ by simply interchanging the roles of $\mathbf{x}_t^c$ and $\mathbf{y}_t^c$.

\subsection{Overall Algorithm for Secure Communication}
By leveraging the detailed solutions for subproblems \eqref{51} and \eqref{62}, we construct a comprehensive AO algorithm to solve problem \eqref{50}. The overall method is presented in Algorithm 4, where $\delta_5$ denotes the convergence threshold for $R_\text{sec}({\tilde{\mathbf{t}}}_c,\bm{\rho}_e,\boldsymbol{\omega}^*)$. The two subproblems \eqref{64} and \eqref{73} are solved iteratively until the increase in  $R_\text{sec}({\tilde{\mathbf{t}}}_c,\bm{\rho}_e,\boldsymbol{\omega}^*)$ falls below $\delta_5$. The total computational complexity of Algorithm 4 is given by $\mathcal{O}(L_t(L_x+L_y)((M_s+N)J(N^{3}+FN^2)+N^{3.5} \ln(1/\kappa))+BNJ(N^{3}+FN^2))$, where $L_t$ denotes the maximum number of iterations for the main loop (steps 7-10), and $L_y$ denotes the maximum number of iterations for optimizing $\mathbf{y}^c_t$ in step 8. Algorithm 4 converges rapidly in practice, as shown in Section VI-A. In practical implementations, a multi-timescale update mechanism can be adopted to reduce the need for frequent antenna movement and to accommodate the distinct temporal characteristics of the wireless channel \cite{RF2-1}. Specifically, since the MAs' positions are primarily determined by slowly varying angular parameters, they can be updated less frequently across multiple transmission intervals. Meanwhile, the beamforming vectors, which must accommodate instantaneous channel variations, can be updated more frequently with low complexity by employing warm-start strategies that exploit the high temporal correlation of the wireless channels.

\begin{algorithm}[ht]
\caption{Overall Optimization Algorithm for Secure Communication}
\label{alg4}
\textbf{Input:} $N$, $D$, $\mathcal{C}_t$, $\delta_5$, $\mathbf{x}^s_t$, $\mathbf{y}^s_t$. \\
\textbf{Output:} $\mathbf{x}_t^c$, $\mathbf{y}_t^c$, $\boldsymbol{\omega}$.
\begin{algorithmic}[1]
    \State Initialization: $\mathbf{x}^{c\;(0)}_t\leftarrow\mathbf{x}^s_t$, $\mathbf{y}^{c\;(0)}_t\leftarrow\mathbf{y}^s_t$.
    \For{$ b = 1 : 1 : B $} 
    \State \parbox[t]{0.4\textwidth}{Based on the optimized MAs' positions from Algorithm 1, the eavesdropper’s AoD information $\hat{\bm{\rho}}_e(b)$ is estimated by \eqref{15} for each noise realization.} 
    \State \parbox[t]{0.4\textwidth}{Optimize $\boldsymbol{\omega}$ according to Algorithm 2 and obtain the secrecy rate $R_\text{sec}(\tilde{\mathbf{t}}_c,\bm{\rho}_e,\boldsymbol{\omega}^*)$.}
    \EndFor
    \State $\hat{\bm{\rho}}_e=\arg\min_{\{\hat{\bm{\rho}}_e(1),\dots,\hat{\bm{\rho}}_e(B)\}}R_\text{sec}(\tilde{\mathbf{t}}_c,\bm{\rho}_e,\boldsymbol{\omega}^*)$
    \While{Increase of $R_\text{sec}(\tilde{\mathbf{t}}_c,\bm{\rho}_e,\boldsymbol{\omega}^*)$  is above $\delta_5$}
       \State Given $\mathbf{y}_t^c$, solve problem \eqref{64} to update $\mathbf{x}_t^c$.
       \State Given $\mathbf{x}_t^c$, solve problem \eqref{73} to update $\mathbf{y}_t^c$.
    \EndWhile
\end{algorithmic}
\end{algorithm}

\section{Simulation Results and Discussions}
In this section, we present simulation results to evaluate the performance of the proposed sensing-assisted secure communication scheme. We consider an ISAC BS equipped with $N=16$ transmit MAs and $M=16$ receive MAs, operating at a carrier wavelength of $\lambda=0.05$ m. The local coordinate origins for the transmit and receive MA regions, $O_t$ and $O_r$, are located at (0, $10\lambda$, 0) m and (0, 0, 0) m in a global 3D Cartesian coordinate system, respectively. The square movement region for both MA arrays has a side length of $A=5\lambda$, and the minimum inter-MA distance is set to $D=\lambda/2$. The number of sensing snapshots is $T=16$, and the RCS is $\epsilon=10$ dBsm. Unless otherwise specified, the sensing transmit power is $P_s=30$ dBm, the maximum communication transmit power is $P_t=20$ dBm, and the noise power is $\sigma_s^2=\sigma_c^2=\sigma_e^2=-90$ dBm. The legitimate receiver is located at a distance of $d_{bc}=70$ m, with AoDs $\theta_c=120^\circ$ and $\varphi_c=90^\circ$, corresponding to spatial AoDs $\alpha_c=0$ and $\beta_c=-0.5$. The eavesdropper is at a distance of $d_{be}=70$ m, with AoDs $\theta_e=120^\circ$ and $\varphi_e=120^\circ$, which yields spatial AoDs $\alpha_e=-0.43$ and $\beta_e=-0.5$. The maximum allowable CRB threshold is set to $\eta=0.001$ to provide a reliable sensing performance for the robust beamforming design.

\begin{figure}[!t]
    \centering
    \begin{minipage}{0.24\textwidth}
        \centering
        \includegraphics[width=\textwidth]{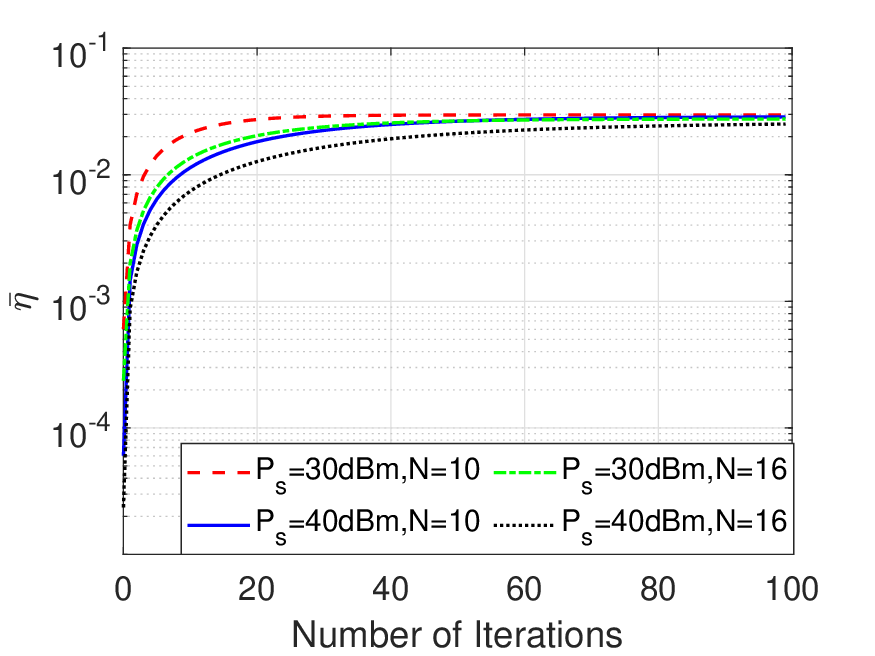}
        \par\smallskip
        (a) Algorithm 1
        \label{fig:sub11}
    \end{minipage}
    \hfill
    \begin{minipage}{0.24\textwidth}
        \centering
        \includegraphics[width=\textwidth]{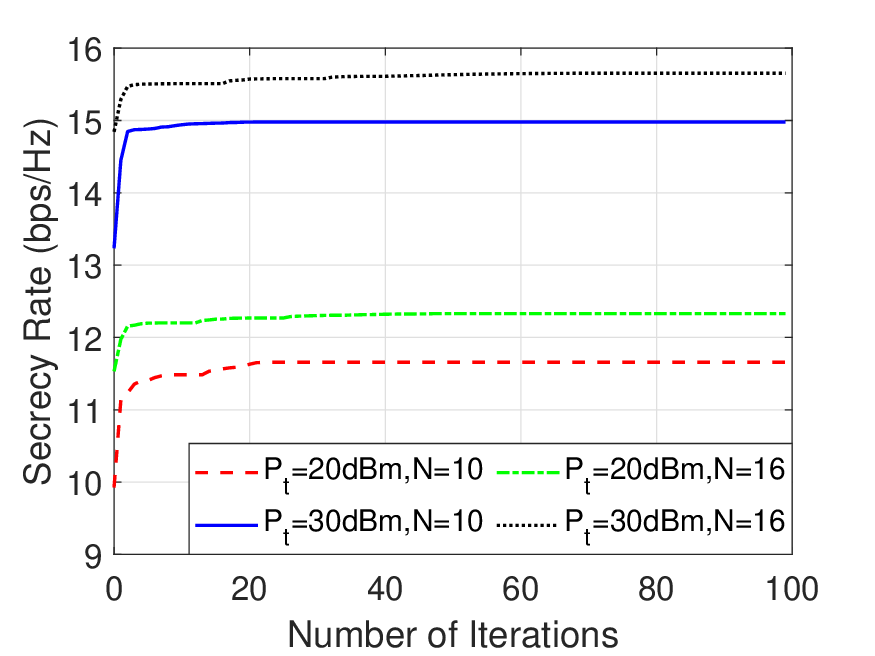}
        \par\smallskip
        (b) Algorithm 4
        \label{fig:sub22}
    \end{minipage}
    \caption{Convergence behavior of the proposed algorithms.}
    \label{fig:add1}
\end{figure}

\subsection{Convergence Behaviour of Proposed Algorithms}
First, we evaluate the convergence behavior of the proposed AO-based algorithms to assess their computational efficiency. Fig. 3 illustrates the convergence trajectories of the objective functions in both the eavesdropper sensing stage (Algorithm 1) and the secure communication stage (Algorithm 4). Specifically, Fig. 3(a) shows the evolution of the objective function $\bar{\eta}$ in Algorithm 1 versus the number of iterations. Under different numbers of MAs and sensing power levels, $\bar{\eta}$ increases monotonically and converges to a stable value within approximately 50 iterations. Fig. 3(b) depicts the convergence behavior of Algorithm 4 in terms of the secrecy rate. For different numbers of MAs and maximum communication power levels, the secrecy rate also increases monotonically and stabilizes within approximately 40 iterations. These results demonstrate that the proposed AO-based algorithms converge reliably and rapidly with moderate computational complexity, validating their practical effectiveness and computational efficiency.

\begin{figure}[!t]
    \centering
    \begin{minipage}{0.45\textwidth}
        \centering
        \includegraphics[width=\textwidth]{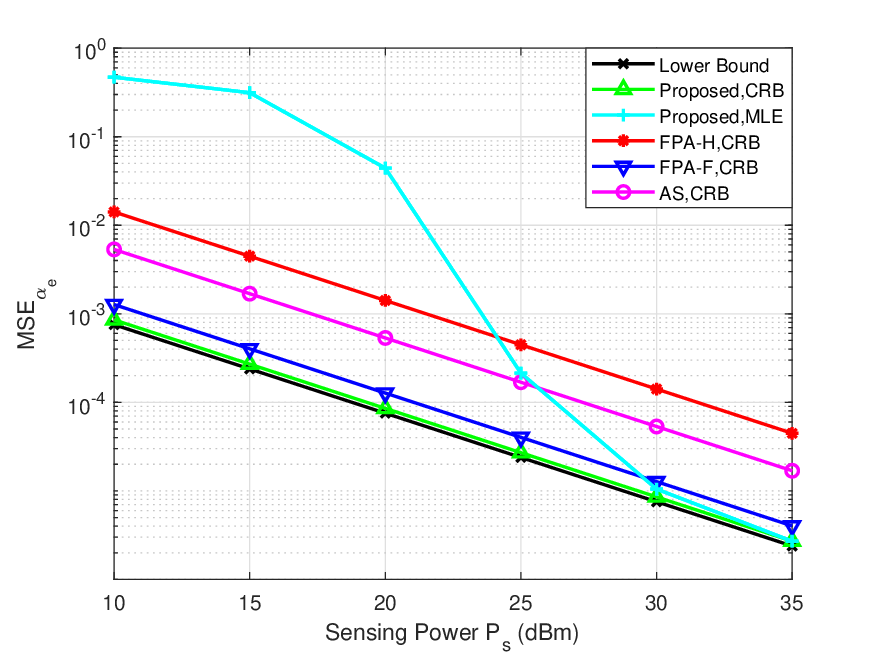}
        \par\smallskip
        (a) $\text{MSE}_{\alpha_e}$
        \label{fig:sub7}
    \end{minipage}
    \hfill
    \begin{minipage}{0.45\textwidth}
        \centering
        \includegraphics[width=\textwidth]{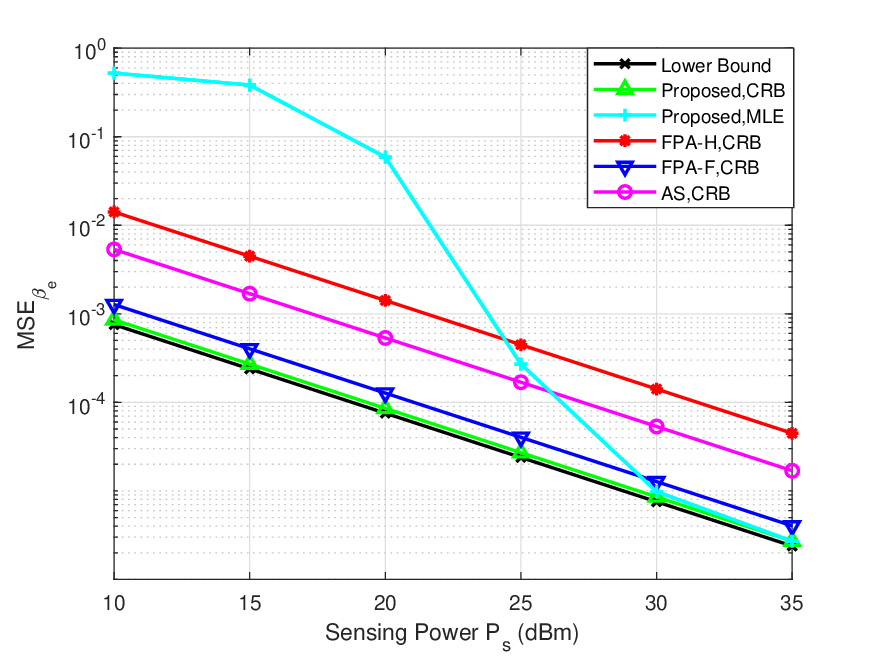}
        \par\smallskip
        (b) $\text{MSE}_{\beta_e}$
        \label{fig:sub8}
    \end{minipage}
    \caption{MSE of eavesdropper AoD estimation versus sensing power.}
    \label{fig:B}
\end{figure}

\subsection{Sensing Performance Comparison}
Then, to evaluate the sensing performance for estimating the eavesdropper’s AoDs, we compare our proposed scheme with three benchmarks: 1) \textbf{FPA-H (Half-wavelength Spacing)}: Transmit and receive uniform planar arrays (UPAs) with $N$ and $M$ antennas, respectively, spaced at $\lambda/2$ intervals in both dimensions. 2) \textbf{FPA-F (Full Aperture)}: Transmit and receive UPAs with antenna spacings of $\frac{A}{\sqrt{N}-1}$ and $\frac{A}{\sqrt{M}-1}$ to maximize the array aperture. 3) \textbf{AS (Antenna Selection)}: An optimal subset of $N$ transmit and $M$ receive antennas is selected from larger UPAs (with $2N$ and $2M$ elements at $\lambda/2$ spacing) via an exhaustive search to minimize the CRB \cite{RF2-12}. We also include the theoretical CRB lower bound for a square region, which is $\frac{\lambda^2\sigma^2_s}{4MP_sT\pi^2A^2|\zeta_s|^2}$ \cite{RF2-13}.

Fig. 4 shows the eavesdropper AoD estimation performance for $\alpha_e$ and $\beta_e$ in terms of both the CRB and the empirical MSE versus the sensing power $P_s$. To numerically validate the derived CRB, the empirical MSE of the MLE obtained via Monte Carlo simulations is also presented. As observed in Figs. 4(a) and 4(b), the proposed scheme achieves significantly lower CRBs than the FPA-H, FPA-F, and AS benchmarks, and its performance closely approaches the theoretical lower bound. This result verifies that jointly optimizing the transmit and receive MAs' positions is highly effective in enhancing AoD estimation precision. Furthermore, in the high sensing power regime (e.g., $P_s \geq 30$ dBm), the MSE of the MLE asymptotically approaches the corresponding CRB, indicating that the derived CRB provides a tight lower bound under moderate-to-high sensing SNR conditions. In contrast, in the low sensing SNR regime, a noticeable gap between the MLE performance and the CRB is observed, which is consistent with the well-known threshold effect in nonlinear parameter estimation. The consistent trends observed for both $\alpha_e$ and $\beta_e$ further validate the effectiveness of the proposed joint MA position optimization.

Fig. 5 illustrates the optimized positions of the transmit and receive MAs during the eavesdropper sensing stage. The positions achieve a balance between dispersion and symmetry with respect to both the horizontal and vertical axes. This configuration effectively maximizes the variance terms while minimizing the covariance terms in the CRB expressions, leading to minimal estimation error. Furthermore, the MAs are positioned near the boundaries of their regions, maximizing the array aperture to improve angular resolution.

\begin{figure}[!t]
    \centering
    \begin{minipage}{0.24\textwidth}
        \centering
        \includegraphics[width=\textwidth]{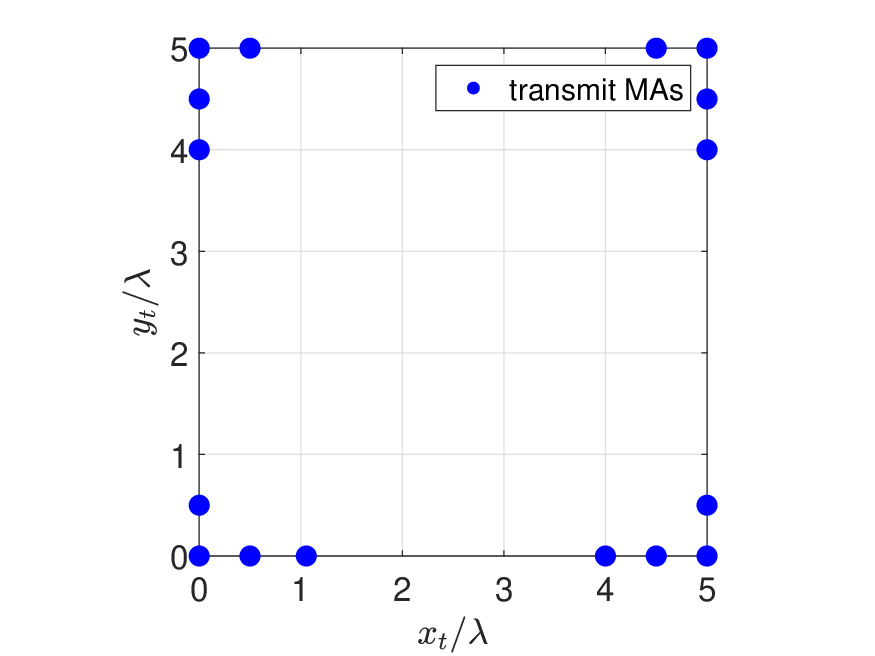}
        \par\smallskip
        (a) Deployed positions of transmit MAs.
        \label{fig:sub5}
    \end{minipage}
    \hfill
    \begin{minipage}{0.24\textwidth}
        \centering
        \includegraphics[width=\textwidth]{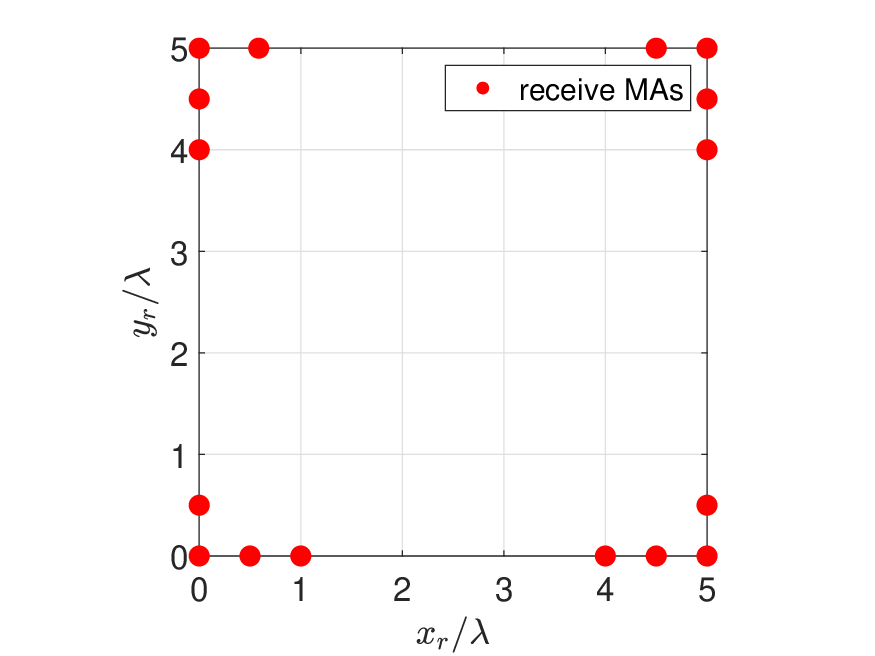}
        \par\smallskip
        (b) Deployed positions of receive MAs.
        \label{fig:sub6}
    \end{minipage}
    \caption{Illustration of the MAs' positions in the eavesdropper sensing stage.}
    \label{fig:A}
\end{figure}

\subsection{Communication Performance Comparison}
Next, we evaluate the secure communication performance of the proposed scheme against several benchmarks: 1) \textbf{MRT (Maximum Ratio Transmission)}: In this scheme, the transmit MAs’ positions are first optimized to maximize the legitimate channel gain. Based on the resulting optimized antenna configuration, the beamforming vector is designed following the MRT principle for the legitimate receiver’s channel, without specifically suppressing information leakage to the eavesdropper. 2) \textbf{MRT-ZF (MRT with Zero-Forcing)}: Similar to the MRT scheme, the transmit MAs’ positions are first optimized to enhance the legitimate channel. Then, the beamforming vector for the information signal is designed via MRT, while the artificial noise (AN) precoding vector is projected onto the null space of the legitimate receiver’s channel to degrade the eavesdropper’s reception without interfering with the legitimate receiver \cite{RF2-C}. The total transmit power is split equally between the information signal and the AN. This equal power allocation is adopted as a representative setting for the MRT-ZF benchmark. 3) \textbf{Ideal Estimation}: This scheme serves as a performance upper bound. It employs the same optimization algorithm as our proposed scheme but assumes perfect, error-free knowledge of the eavesdropper's true AoDs. 4) \textbf{Estimated-as-True}: In this non-robust scheme, the transmit MA positions and beamforming vectors are optimized by treating the estimated eavesdropper AoDs, $\hat{\theta}_e$ and $\hat{\varphi}_e$, without accounting for the uncertainty region $\Xi_2$. In this case, $\Xi_2$ collapses to a singleton set containing only the estimated AoDs. 5) \textbf{FPA-H}: In this scheme, the transmit and receive antennas are arranged as UPAs with half-wavelength spacing both vertically and horizontally, consistent with the sensing benchmark scheme. Both the eavesdropper sensing and secure communication stages are performed under this fixed antenna configuration, and the robust beamforming vectors are optimized within $\Xi_2$.

\begin{figure}[!t]
    \centering  
    \includegraphics[width=0.45\textwidth]{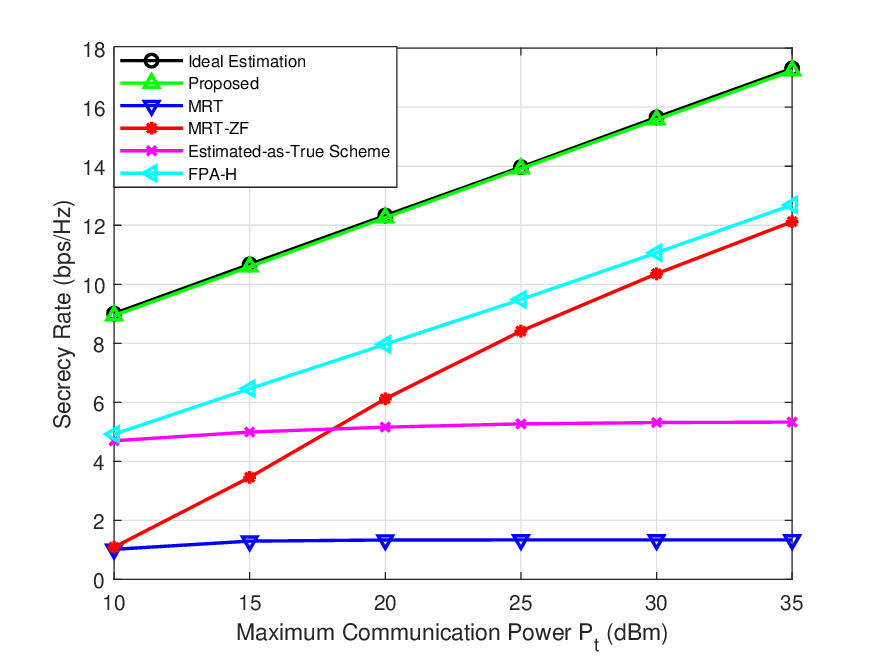} 
    \caption{Secrecy rates of different schemes versus maximum communication power.}
    \label{fig:C}  
\end{figure}

Fig. 6 shows the secrecy rate of different schemes versus the maximum communication power $P_t$. The secrecy rate of the proposed scheme is very close to that of the Ideal Estimation scheme, and both significantly outperform all other benchmarks. Specifically, the noticeable performance gap between the proposed scheme and the FPA-H benchmark clearly demonstrates the secrecy rate gains enabled by the spatial flexibility and additional spatial DoFs provided by MAs. Furthermore, the superiority of the proposed scheme over the MRT, MRT-ZF, and Estimated-as-True schemes highlights the critical importance of accurately sensing the eavesdropper’s direction and employing a robust beamforming design that accounts for estimation uncertainty. In addition, while the secrecy rates of the Ideal, Proposed, FPA-H, and MRT-ZF schemes increase with $P_t$, the MRT and Estimated-as-True schemes exhibit negligible improvement. This is because the MRT scheme lacks any anti-eavesdropping mechanism, and the Estimated-as-True scheme is highly sensitive to estimation errors, causing the eavesdropping rate to grow with $P_t$, thus negating any potential gains in secrecy rate.

\begin{figure}[!t]
    \centering
    \begin{minipage}{0.24\textwidth}
        \centering
        \includegraphics[width=\textwidth]{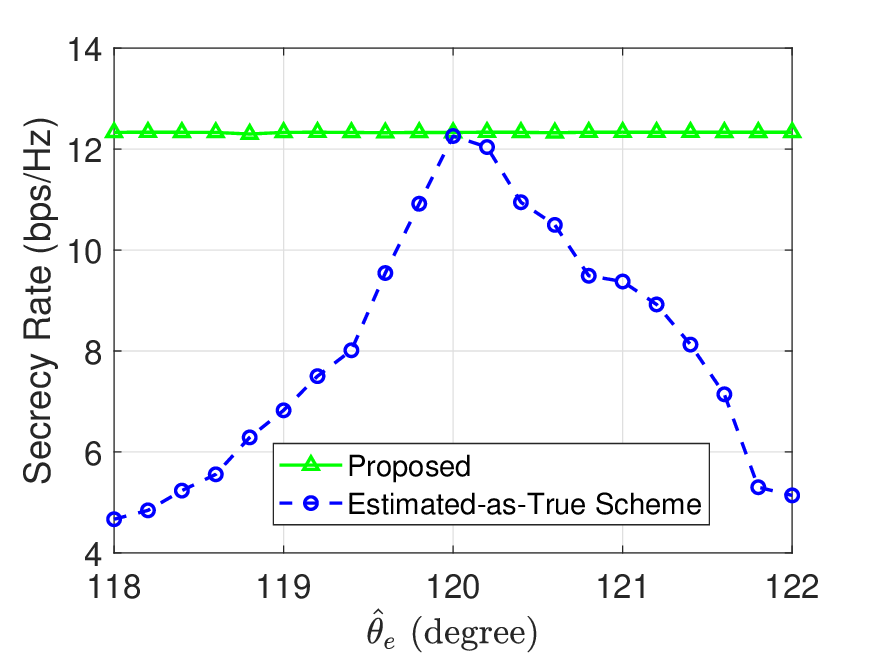}
        \par\smallskip
        (a) Secrecy rate versus estimated $\hat{\theta}_e$ for $\hat{\varphi}_e=120^\circ$.
        \label{fig:sub9}
    \end{minipage}
    \hfill
    \begin{minipage}{0.24\textwidth}
        \centering
        \includegraphics[width=\textwidth]{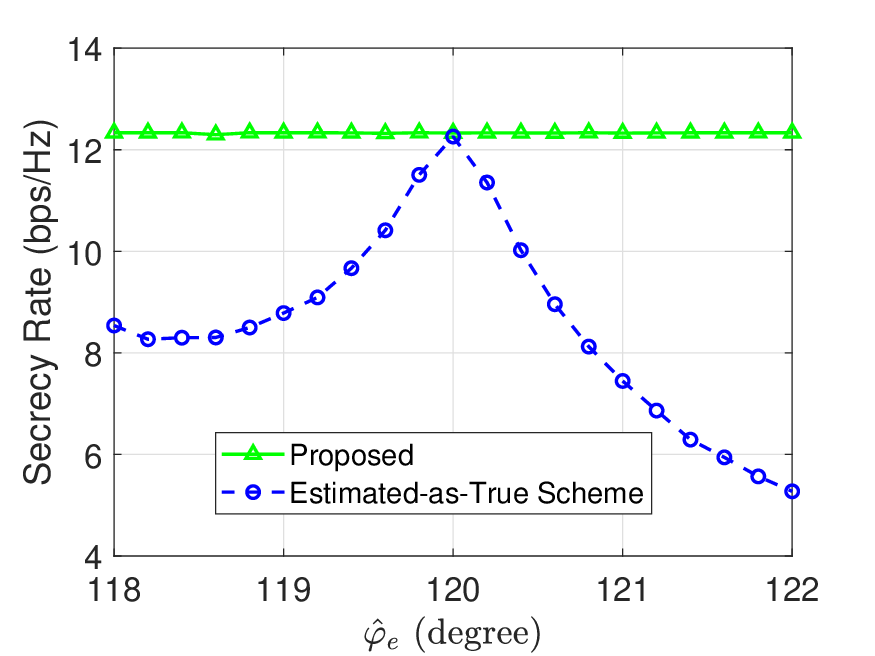}
        \par\smallskip
        (b) Secrecy rate versus estimated $\hat{\varphi}_e$ for $\hat{\theta}_e=120^\circ$.
        \label{fig:sub10}
    \end{minipage}
    \caption{Secrecy rates of the proposed and Estimated-as-True schemes versus estimated AoDs of the eavesdropper.}
    \label{fig:E}
\end{figure}

\begin{figure}[!t]
    \centering  
    \includegraphics[width=0.45\textwidth]{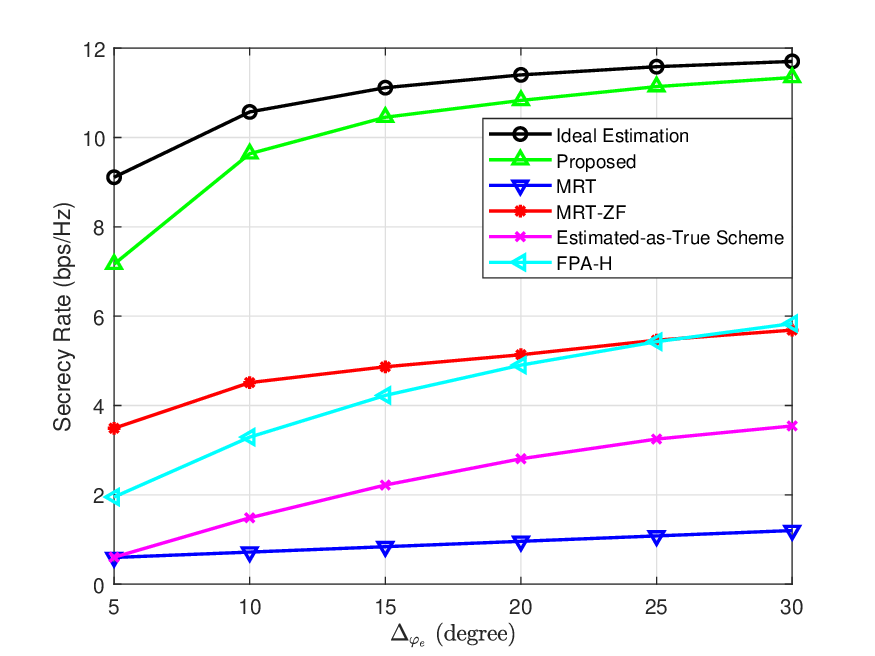}  
    \caption{Secrecy rates of different scheme versus width of the eavesdropper’s azimuth AoD region.}
    \label{fig:D}  
\end{figure} 

Fig. 7 highlights the robustness of the proposed scheme compared to the non-robust Estimated-as-True scheme. The true AoDs are fixed at $\theta_e=120^\circ$ and $\varphi_e=120^\circ$, while the estimated AoDs vary within the uncertainty interval $[118^\circ, 122^\circ]$. As shown, the proposed scheme maintains a consistently high secrecy rate across the entire range, whereas the performance of the Estimated-as-True scheme degrades sharply as the estimated AoDs deviate from the true values. This confirms that by optimizing for the worst-case scenario within the AoD uncertainty region, our proposed scheme achieves superior robustness and effectively suppresses eavesdropping, regardless of the specific estimation outcome.

To further investigate the robustness of different schemes against spatial location variations, we evaluate the average secrecy rate versus the width of the eavesdropper’s azimuth AoD region, denoted by $\Delta_{\varphi_e}$. Specifically, the legitimate receiver's AoDs are fixed, while the eavesdropper's azimuth AoD is randomly generated following a uniform distribution within a predefined angular region centered around the legitimate receiver, i.e., $\varphi_e \sim \mathcal{U}(\varphi_b-\Delta_{\varphi_e}, \varphi_b+\Delta_{\varphi_e})$, with $d_{be}=d_{bc}$. The secrecy rate is averaged over multiple independent random realizations of $\varphi_e$. Fig. 8 depicts the secrecy rate versus $\Delta_{\varphi_e}$ for different schemes. It is observed that the secrecy rates of all schemes increase monotonically with $\Delta_{\varphi_e}$. This is because a wider uncertainty region increases the probability of a larger angular separation between the eavesdropper and the legitimate receiver. This, in turn, reduces their spatial channel correlation, thereby facilitating more effective spatial nulling and information leakage suppression at the BS. Moreover, the secrecy rate of the proposed scheme closely approaches that of the Ideal Estimation scheme, and both significantly outperform all other benchmarks. These results clearly demonstrate the superior robustness of the proposed joint MA and beamforming design against spatial variations of the eavesdropper.

\begin{figure}[!t]
    \centering  
    \includegraphics[width=0.45\textwidth]{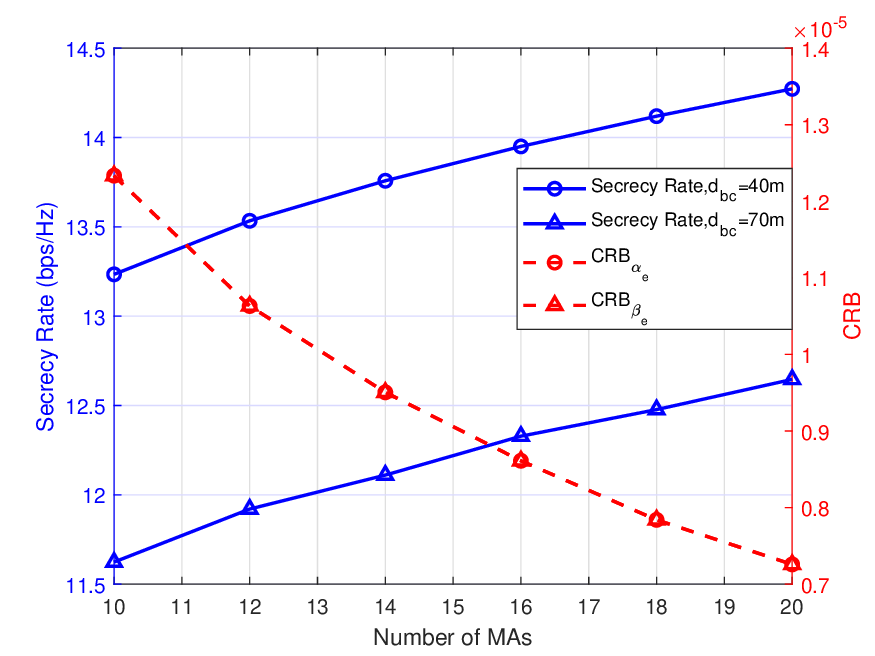}  
    \caption{Secrecy rate and CRB of the proposed scheme versus number of MAs.}
    \label{fig:11}  
\end{figure} 

\begin{figure}[!t]
    \centering  
    \includegraphics[width=0.45\textwidth]{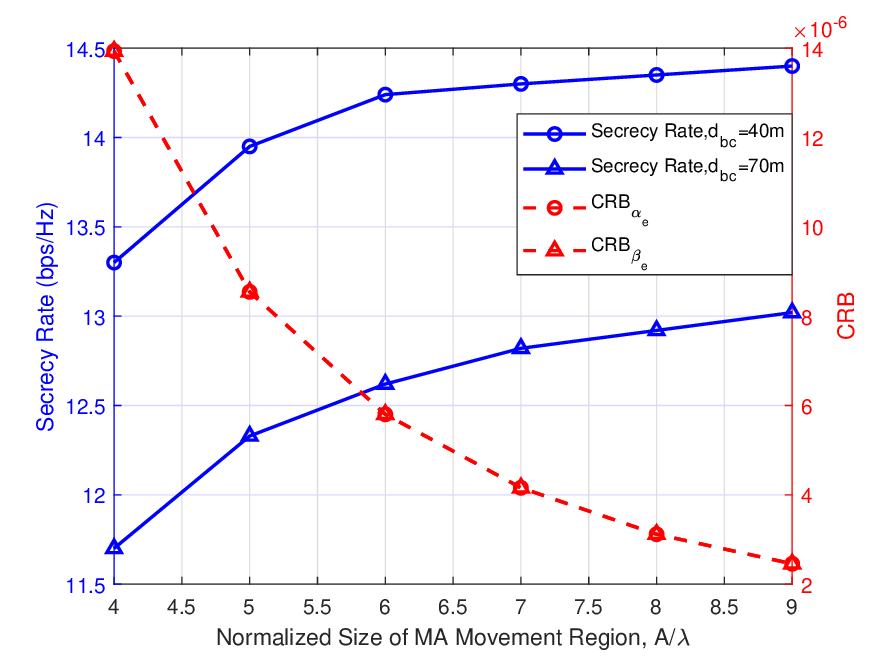}  
    \caption{Secrecy rate and CRB of the proposed scheme versus normalize size of MA movement region.}
    \label{fig:12}  
\end{figure}

\subsection{Impact of MA Parameters on Joint Sensing and Communication Performance}
Finally, we evaluate the impact of key MA parameters, including the number of MAs and the size of MA movement region, on both sensing and communication performance.

Fig. 9 illustrates the secrecy rate under different legitimate receiver distances and the corresponding CRB of the proposed scheme versus the number of MAs. It is observed that as the number of MAs increases, the secrecy rate for both $d_{bc}=40$ m and $d_{bc}=70$ m increases steadily, while the CRBs for both $\alpha_e$ and $\beta_e$ decrease accordingly. This joint performance improvement is attributed to the increased spatial DoFs provided by additional MAs, which not only facilitate more accurate estimation of the eavesdropper’s AoDs but also enable more flexible spatial beamforming to suppress information leakage. These results demonstrate that increasing the number of MAs is beneficial for enhancing both sensing accuracy and secrecy performance.

Fig. 10 shows the impact of the normalized size of the MA movement region on the system performance. As the MA movement region expands, the CRBs decrease and the secrecy rate improves. This is because enlarging the feasible movement region increases the effective array aperture and provides additional spatial flexibility. Consequently, the MAs can access more favorable spatial positions to enhance AoD estimation accuracy and construct better channel conditions for secure communication. However, the performance improvement gradually saturates as the movement region becomes sufficiently large. This observation reveals that a moderate-size movement region is sufficient to capture most of CRB reduction and secrecy rate improvement, providing a useful guidance for practical MA deployment.

\section{Conclusion}
In this paper, we proposed a sensing-assisted secure communication scheme for an MA-aided ISAC system. To achieve security, we formulated a joint optimization problem that cohesively designs the system's sensing and communication stages. In the eavesdropper sensing stage, we derived the closed-form CRB for the estimated AoDs to characterize the fundamental relationship between MA positions and the AoD estimation range. Subsequently, in the secure communication stage, we designed a robust beamforming vector within this estimation range and further adjusted the transmit MAs' positions to maximize the worst-case secrecy rate. The challenging non-convex subproblems for the two stages were effectively solved using an AO algorithm and a backward induction-based method, respectively. Simulation results validated the effectiveness and robustness of the proposed scheme, demonstrating its superior performance compared to various benchmark schemes. Extending the proposed scheme to more generalized and complex scenarios, such as multi-eavesdropper, multi-cell, and mobile settings, as well as integrating more advanced MA architectures like six-dimensional MA (6DMA) systems with antenna positioning and rotation \cite{RF3-1}-\cite{RF3-4}, represent highly promising and critical research directions to further improve sensing-assisted secure communication performance. 

\appendices

\section{Proof of Lemma 1}
To simplify the derivation, the received signal matrix in \eqref{6} is first vectorized as
\begin{align}\label{74}
\mathbf{\overline{y}}\triangleq\text{vec}\left(\mathbf{Y}_s)=\zeta_s\mathbf{a}(\tilde{\mathbf{t}},\tilde{\mathbf{r}},\bm{\rho}_e\right)+\text{vec}\left(\mathbf{Z}_s\right), 
\end{align}
where $\mathbf{a}(\tilde{\mathbf{t}},\tilde{\mathbf{r}},\bm{\rho}_e)\triangleq\text{vec}\left(\mathbf{f}_e (\tilde{\mathbf{r}},\bm{\rho}_e)\mathbf{g}_e(\tilde{\mathbf{t}},\bm{\rho}_e)^H\mathbf{X}_s\right)$. The MLE of $\zeta_s$ and $\bm{\rho}_e$ can then be found by solving the following problem:
\begin{align}\label{75}
(\hat{\zeta}_s,\hat{\bm{\rho}}_e)=\arg \min_{\bar{\zeta}_s,\bar{\bm{\rho}}_e} \left\|\mathbf{\overline{y}}-\bar{\zeta}_s\mathbf{a}(\tilde{\mathbf{t}},\tilde{\mathbf{r}},\bar{\bm{\rho}}_e)\right\|_2^2.
\end{align}
For any given $\bm{\rho}_e$, the minimization problem in \eqref{75} w.r.t $\bar{\zeta}_s$ is a standard linear least squares problem, which admits the optimal solution $\hat{\zeta}_s=\frac{\mathbf{a}(\tilde{\mathbf{t}},\tilde{\mathbf{r}},{\bm{\rho}}_e)^H\mathbf{\overline{y}}}{\left\|\mathbf{a}(\tilde{\mathbf{t}},\tilde{\mathbf{r}},{\bm{\rho}}_e)\right\|_2^2}$. By substituting it into \eqref{75}, we can obtain
\begin{align}\label{77}
\left\|\mathbf{\overline{y}}-\hat{\zeta}_s\mathbf{a}(\tilde{\mathbf{t}},\tilde{\mathbf{r}},{\bm{\rho}}_e)\right\|_2^2=\left\|\mathbf{\overline{y}}\right\|^2_2-\frac{\left|\mathbf{\overline{y}}^H\mathbf{a}(\tilde{\mathbf{t}},\tilde{\mathbf{r}},{\bm{\rho}}_e)\right|^2}{\left\|\mathbf{a}(\tilde{\mathbf{t}},\tilde{\mathbf{r}},{\bm{\rho}}_e)\right\|_2^2}.
\end{align}
To simplify this expression, we first analyze the numerator term. Using the properties of the vectorization operator, we can rewrite it as
\begin{align}\label{78}
\left|\mathbf{\overline{y}}^H\mathbf{a}(\tilde{\mathbf{t}},\tilde{\mathbf{r}},{\bm{\rho}}_e)\right|^2=\left|(\mathbf{f}_e (\tilde{\mathbf{r}},\bm{\rho}_e)\otimes\mathbf{g}_e(\tilde{\mathbf{t}},\bm{\rho}_e))^H\text{vec}\left(\mathbf{X}_s(\mathbf{Y}_s)^H\right)\right|^2.
\end{align}
Next, we simplify the denominator term:
\begin{align}\label{79}
\left\|\mathbf{a}(\tilde{\mathbf{t}},\tilde{\mathbf{r}},{\bm{\rho}}_e)\right\|_2^2=MTP_s.
\end{align}
Since both $\left\|\mathbf{\overline{y}}\right\|^2_2$ and $\left\|\mathbf{a}(\tilde{\mathbf{t}},\tilde{\mathbf{r}},{\bm{\rho}}_e)\right\|_2^2$ are constant w.r.t. ${\bm{\rho}}_e$, minimizing the expression in \eqref{77} is equivalent to maximizing the numerator term. Thus, the MLE of ${\bm{\rho}}_e$ is given by
\begin{align}\label{80}
\hat{\bm{\rho}}_e=\arg \max_{\bar{\bm{\rho}}_e} \left|(\mathbf{f}_e (\tilde{\mathbf{r}},\bm{\rho}_e)\otimes\mathbf{g}_e(\tilde{\mathbf{t}},\bm{\rho}_e))^H\text{vec}\left(\mathbf{X}_s(\mathbf{Y}_s)^H\right)\right|^2.
\end{align}
This completes the proof.

\section{Proof of Lemma 2}
We define $\bm{\zeta}\triangleq\left[\Re\left(\zeta_s\right),\Im\left(\zeta_s\right)\right]^T$ to comprise the real and imaginary parts of the complex channel coefficient. Additionally, we define the vector of unknown parameters as $\bm{\upsilon}\triangleq\left[\bm{\rho}_e,\bm{\zeta}\right]^T$. To facilitate the derivation of the CRB, we further introduce an auxiliary function $\mathbf{q}(\bm{\rho}_e,\bm{\zeta})\triangleq\zeta_s\mathbf{a}(\tilde{\mathbf{t}},\tilde{\mathbf{r}},{\bm{\rho}}_e)$. The FIM for estimating $\bm{\upsilon}$ is denoted as $\mathbf{F}$, which is given by
\begin{align}\label{81}
\mathbf{F} = \begin{bmatrix}
\mathbf{J}_{\bm{\rho}_e,\bm{\rho}_e} & \mathbf{J}_{\bm{\rho}_e,\bm{\zeta}}  \\
\mathbf{J}^T_{\bm{\rho}_e,\bm{\zeta}} & \mathbf{J}_{\bm{\zeta},\bm{\zeta}}  
\end{bmatrix}.
\end{align}
The element in the $i$-th row and $j$-th column ($i, j \in \{1, 2\}$) of $\mathbf{F}$ is calculated by $\mathbf{F}_{i,j} = 2 \Re\left\{ \frac{\partial\mathbf{q}(\bm{\rho}_e,\bm{\zeta})^H}{\partial\bm{\upsilon}_i} \mathbf{R}_z^{-1} \frac{\partial\mathbf{q}(\bm{\rho}_e,\bm{\zeta})^H}{\partial\bm{\upsilon}_j} \right\},$ and the noise covariance $\mathbf{R}_z=\sigma^2_s\mathbf{I}_M$. The partial derivatives of $\mathbf{q}(\bm{\rho}_e,\bm{\zeta})$ w.r.t. $\bm{\rho}_e$ and $\bm{\zeta}$ are
\begin{align}
\frac{\partial\mathbf{q}(\bm{\rho}_e,\bm{\zeta})}{\partial\bm{\rho}_e}&=\zeta_s\begin{bmatrix}
\text{vec}((\dot{\mathbf{f}}_{\alpha_e}\mathbf{g}_e (\tilde{\mathbf{t}},\bm{\rho}_e)^H+\dot{\mathbf{g}}_{\alpha_e}\mathbf{f}_e (\tilde{\mathbf{r}},\bm{\rho}_e)^H)\mathbf{X}_s)\\
\text{vec}((\dot{\mathbf{f}}_{\beta_e}\mathbf{g}_e (\tilde{\mathbf{t}},\bm{\rho}_e)^H+\dot{\mathbf{g}}_{\beta_e}\mathbf{f}_e (\tilde{\mathbf{r}},\bm{\rho}_e)^H)\mathbf{X}_s)
\end{bmatrix}^T,\label{86}\\
\frac{\partial\mathbf{q}(\bm{\rho}_e,\bm{\zeta})}{\partial\bm{\zeta}}&=\mathbf{a}(\tilde{\mathbf{t}},\tilde{\mathbf{r}},{\bm{\rho}}_e)[1,j],\label{87}
\end{align}
where 
$\dot{\mathbf{f}}_{\alpha_e}=\frac{\partial \mathbf{f}_e (\tilde{\mathbf{r}},\bm{\rho}_e)}{\partial \alpha_e}=j\frac{2\pi}{\lambda}\text{diag}(\mathbf{x}_r)\mathbf{f}_e (\tilde{\mathbf{r}},\bm{\rho}_e),\quad\dot{\mathbf{g}}_{\alpha_e}=$ $\frac{\partial \mathbf{g}_e (\tilde{\mathbf{t}},\bm{\rho}_e)}{\partial \alpha_e}=j\frac{2\pi}{\lambda}\text{diag}(\mathbf{x}_t)\mathbf{g}_e (\tilde{\mathbf{t}},\bm{\rho}_e)$,
$\dot{\mathbf{f}}_{\beta_e}=\frac{\partial \mathbf{f}_e (\tilde{\mathbf{r}},\bm{\rho}_e)}{\partial \beta_e}=j\frac{2\pi}{\lambda}\text{diag}(\mathbf{y}_r)\mathbf{f}_e (\tilde{\mathbf{r}},\bm{\rho}_e)$,
$\dot{\mathbf{g}}_{\beta_e}=\frac{\partial \mathbf{g}_e (\tilde{\mathbf{t}},\bm{\rho}_e)}{\partial \beta_e}=j\frac{2\pi}{\lambda}\text{diag}(\mathbf{y}_t)\times$ $\mathbf{g}_e (\tilde{\mathbf{t}},\bm{\rho}_e)$.
After some algebraic manipulations, we obtain the explicit expressions for the FIM blocks:
\begin{align}
\mathbf{J}_{\bm{\rho}_e,\bm{\rho}_e}&=\frac{8P_sT\pi^2|\zeta_s|^2}{N\lambda^2\sigma^2_s}
\begin{bmatrix}
A_1 & A_2\\
A_3 & A_4
\end{bmatrix},\label{92}\\
\mathbf{J}_{\bm{\rho}_e,\bm{\zeta}}&=\frac{4\pi P_sT}{N\lambda^2\sigma_s^2}\begin{bmatrix}
N\sum_{m=1}^Mx^r_m-M\sum_{n=1}^Nx^r_n \\
N\sum_{m=1}^My^r_m-M\sum_{n=1}^Ny^r_n 
\end{bmatrix},\label{93}\\
\mathbf{J}_{\bm{\zeta},\bm{\zeta}}&=\frac{2P_sTM}{\sigma_s^2}\mathbf{I}_2,\label{94}
\end{align}
where $A_1=N\sum_{m=1}^M(x_m^r)^2-2\sum_{n=1}^Nx^t_n\sum_{m=1}^Mx^r_m+M\sum_{n=1}^N(x_n^t)^2,\quad A_2=N\sum_{m=1}^M(x_m^ry_m^r)+M\sum_{n=1}^N(x_n^ty_n^t)$ $-\sum_{n=1}^Nx^t_n\sum_{m=1}^My^r_m-\sum_{n=1}^Ny^t_n\sum_{m=1}^Mx^r_m,\quad
A_3=N\sum_{m=1}^M(x_m^ry_m^r)+M\sum_{n=1}^N(x_n^ty_n^t)$ $-\sum_{n=1}^Ny^t_n\sum_{m=1}^Mx^r_m-\sum_{n=1}^Nx^t_n\sum_{m=1}^My^r_m,\; A_4=N\sum_{m=1}^M(y_m^r)^2+M\sum_{n=1}^N(y_n^t)^2$ $-2\sum_{n=1}^Ny^t_n\sum_{m=1}^My^r_m$.

The matrix formed by the first two rows and first two columns of the inverse matrix of $\mathbf{F}$ can be written as 
\begin{align}\label{102}
\mathbf{Q}&\triangleq\begin{bmatrix}
\mathbf{F}^{-1}[1,1] & \mathbf{F}^{-1}[1,2]  \\
\mathbf{F}^{-1}[2,1] & \mathbf{F}^{-1}[2,2]  
\end{bmatrix}=\left[\mathbf{J}_{\bm{\rho}_e,\bm{\rho}_e}-\mathbf{J}_{\bm{\rho}_e,\bm{\zeta}}\mathbf{J}^{-1}_{\bm{\zeta},\bm{\zeta}}\mathbf{J}^T_{\bm{\rho}_e,\bm{\zeta}}\right]^{-1}\nonumber\\
&=\begin{bmatrix}
v(\mathbf{y}_t)+v(\mathbf{y}_r) & -c(\mathbf{x}_t,\mathbf{y}_t)-c(\mathbf{x}_r,\mathbf{y}_r)  \\
-c(\mathbf{x}_t,\mathbf{y}_t)-c(\mathbf{x}_r,\mathbf{y}_r) & v(\mathbf{x}_t)+v(\mathbf{x}_r)  
\end{bmatrix}\times\nonumber\\
&\frac{G}{\left(v(\mathbf{x}_t)+v(\mathbf{x}_r)\right)\left(v(\mathbf{y}_t)+v(\mathbf{y}_r)\right)-(c(\mathbf{x}_t,\mathbf{y}_t)+c(\mathbf{x}_r,\mathbf{y}_r))^2},
\end{align}
where $G=\frac{\lambda^2\sigma^2_s}{8MP_sT\pi^2|\zeta_s|^2}$. According to the definition of FIM, the CRBs for estimated AoDs are given by
\begin{align}
\text{CRB}_{\alpha_e}(\tilde{\mathbf{t}},\tilde{\mathbf{r}})&=\mathbf{Q}[1,1]=\frac{G}{v(\mathbf{x}_t)+v(\mathbf{x}_r)-\frac{(c(\mathbf{x}_t,\mathbf{y}_t)+c(\mathbf{x}_r,\mathbf{y}_r))^2}{v(\mathbf{y}_t)+v(\mathbf{y}_r)}},\label{103}\\
\text{CRB}_{\beta_e}(\tilde{\mathbf{t}},\tilde{\mathbf{r}})&=\mathbf{Q}[2,2]=\frac{G}{v(\mathbf{y}_t)+v(\mathbf{y}_r)-\frac{(c(\mathbf{x}_t,\mathbf{y}_t)+c(\mathbf{x}_r,\mathbf{y}_r))^2}{v(\mathbf{x}_t)+v(\mathbf{x}_r)}}.\label{104}
\end{align}
This completes the proof of Lemma 2.

\end{document}